\documentclass[aps,prx,twocolumn,showpacs,amsmath,amssymb,floatfix,superscriptaddress,citeautoscript]{revtex4-1}

\usepackage{amsmath, amsthm, amssymb}
\usepackage{amsfonts}
\usepackage{graphicx}
\usepackage{subfigure}
\usepackage{color}
\usepackage{times}
\usepackage{float}
\usepackage{natbib}
\usepackage{hyperref}
\usepackage{verbatim}

\hypersetup{
        colorlinks=true,
}

\newcommand{\be}{\begin{equation} }
\newcommand{\ee}{\end{equation} }
\newcommand{\ba}{\begin{eqnarray} }
\newcommand{\ea}{\end{eqnarray} }

\makeatletter

\@ifundefined{textcolor}{}
{%
 \definecolor{BLACK}{gray}{0}
 \definecolor{WHITE}{gray}{1}
 \definecolor{RED}{rgb}{1,0,0}
 \definecolor{GREEN}{rgb}{0,1,0}
 \definecolor{BLUE}{rgb}{0,0,1}
 \definecolor{CYAN}{cmyk}{1,0,0,0}
 \definecolor{MAGENTA}{cmyk}{0,1,0,0}
 \definecolor{YELLOW}{cmyk}{0,0,1,0}
}

\makeatother

\begin{document}

\title{Topological superconductivity in a multichannel Yu-Shiba-Rusinov chain}

\author{Junhua Zhang}
\affiliation{Department of Physics, College of William and Mary, Williamsburg, Virginia 23187, USA}

\author{Younghyun Kim}
\affiliation{Department of Physics, University of California,  Santa Barbara, California 93106, USA}

\author{E. Rossi}
\affiliation{Department of Physics, College of William and Mary, Williamsburg, Virginia 23187, USA}

\author{Roman M. Lutchyn}
\affiliation{Station Q, Microsoft Research, Santa Barbara, California 93106-6105, USA}

\date{\today}

\begin{abstract}
Chains of magnetic atoms placed on the surface of an s-wave superconductor with large spin-orbit coupling provide a promising platform for the realization
of topological superconducting states characterized by the presence of Majorana zero-energy modes.
In this work we study the properties of one-dimensional chains of Yu-Shiba-Rusinov states induced by magnetic impurities
using a realistic model for the magnetic atoms that includes the presence of multiple scattering channels.
These channels are mixed by spin-orbit coupling and, via the hybridization  of the Yu-Shiba-Rusinov states at different sites of the chain,
result in a multi--band structure  for the chain.
We obtain the topological phase diagram for such band structure
and show that the inclusion of higher bands can greatly enlarge the phase space for the realization of topological states.
\end{abstract}

\pacs{
73.20.Hb,
74.78.-w, 
75.70.Tj, 
}

\maketitle

\section{introduction}

The search for Majorana zero modes (Majoranas) in condensed matter
systems has been an active and exciting pursuit~\cite{Reich,
Brouwer_Science, Wilczek2012, LeeYazdani}. The reason for much
excitement
is due to the theoretical prediction that these
modes manifest non-Abelian quantum
statistics\cite{Moore1991,Nayak1996,ReadGreen, Ivanov}, and, as such,
would open the possibility to realize {\em topological quantum computing} ~\cite{TQCreview,
AliceaRev, BeenakkerReview, NayakReview2015}.
Currently, errors caused by the decoherence of the quantum states used to
encode the data constitute the biggest fundamental obstacle for the realization
of a scalable quantum computer.
In topological quantum computing this obstacle is overcome by the
topological protection of the quantum states used to encode the
information.
Most platforms for realizing topological phases of matter
supporting Ising anyons (i.e. exotic defects binding Majorana zero
modes) involve superconducting heterostructures~\cite{FuKane, Fu&Kane09, Sau2010,
Alicea10, Lutchyn2010, Oreg2010, Duckheim'11, SB'11, Flensberg'12,
Potter'12, Martin'12, Mourik2012, Rokhinson2012, Das2012, Deng2012,
Fink2012, Churchill2013, Deng_arxiv2014, Krogstrup2015, Chang2015,
Higginbotham2015}.
Recently, several works
~\cite{Choy'11,
nadj2013, klinovaja2013,braunecker2013,vazifeh2013, pientka2013,
Pientka2013b, poyhonen2014, kim2014, brydon2014, Ebisu14, LiJ2014, weststrom2015, Peng15, Heimes15,
Rontynen2014, LiJ15, Sau2015}
have proposed that a chain of magnetic atoms
placed on the surface of a superconductor
can be in a robust topological phase characterized by the presence
of Majorana modes located at its ends.
In addition, recent experimental results~\cite{nadjperge2014}
have shown that a chain of Fe atoms placed on the surface of
superconducting Pb exhibits a zero-bias peak localized at its ends,
consistent with the presence of Majorana modes.
A very recent preprint also presents experimental results for a similar system~\cite{pawlak2015}.
A full understanding of the experiment presented
in Refs.~\onlinecite{nadjperge2014, pawlak2015} is still being developed,
for example, the height of the zero-bias peak
is a small fraction of the predicted
universal value of $2e^2/h$~\cite{Fidkowski2012, lutchyn_andreev13},
a fact that could be attributed, for example, to finite temperature broadening or disorder.

The potential of a chain of magnetic impurities, placed on the surface of a superconductor,
to be in a robust topological phase, calls for a thorough
theoretical understanding of this system.
So far most of the works have assumed one bound state per magnetic
atom, corresponding to the zeroth angular momentum channel ($l=0$) of
Yu-Shiba-Rusinov(YSR) states~\cite{Yu1965, Shiba68, Rusinov69}.
However, realistic adatoms are expected to induce several
bound states corresponding to different
angular momentum scattering channels (i.e. $l=0,\pm 1, $ etc.).
It has been shown experimentally that partial waves beyond
$s$-wave are essential to explain the energy spectrum of magnetic atoms
such as Mn, Cr and Fe~\cite{Kunz'80, ji2008, Moca08}. Furthermore, the YSR states originating from the $l=0$ channel (s-wave) are
not always the lowest energy eigenstates. It is therefore necessary to
understand the interplay of different angular momentum channels,
in particular in the presence of significant Rashba spin-orbit coupling(SOC)~\cite{kim2014b}.
The Rashba SOC term in the effective low-energy Hamiltonian describing the fermionic degrees of freedom
is expected to be present at the surface due to broken inversion symmetry~\cite{Ast2007}. Moreover, the presence of SOC is required in order to
have a stable topological phase for a ferromagnetically ordered chain~\cite{kim2014, nadjperge2014} placed on the surface of an s-wave superconductor.

\begin{figure}
\includegraphics[width=3.2in]{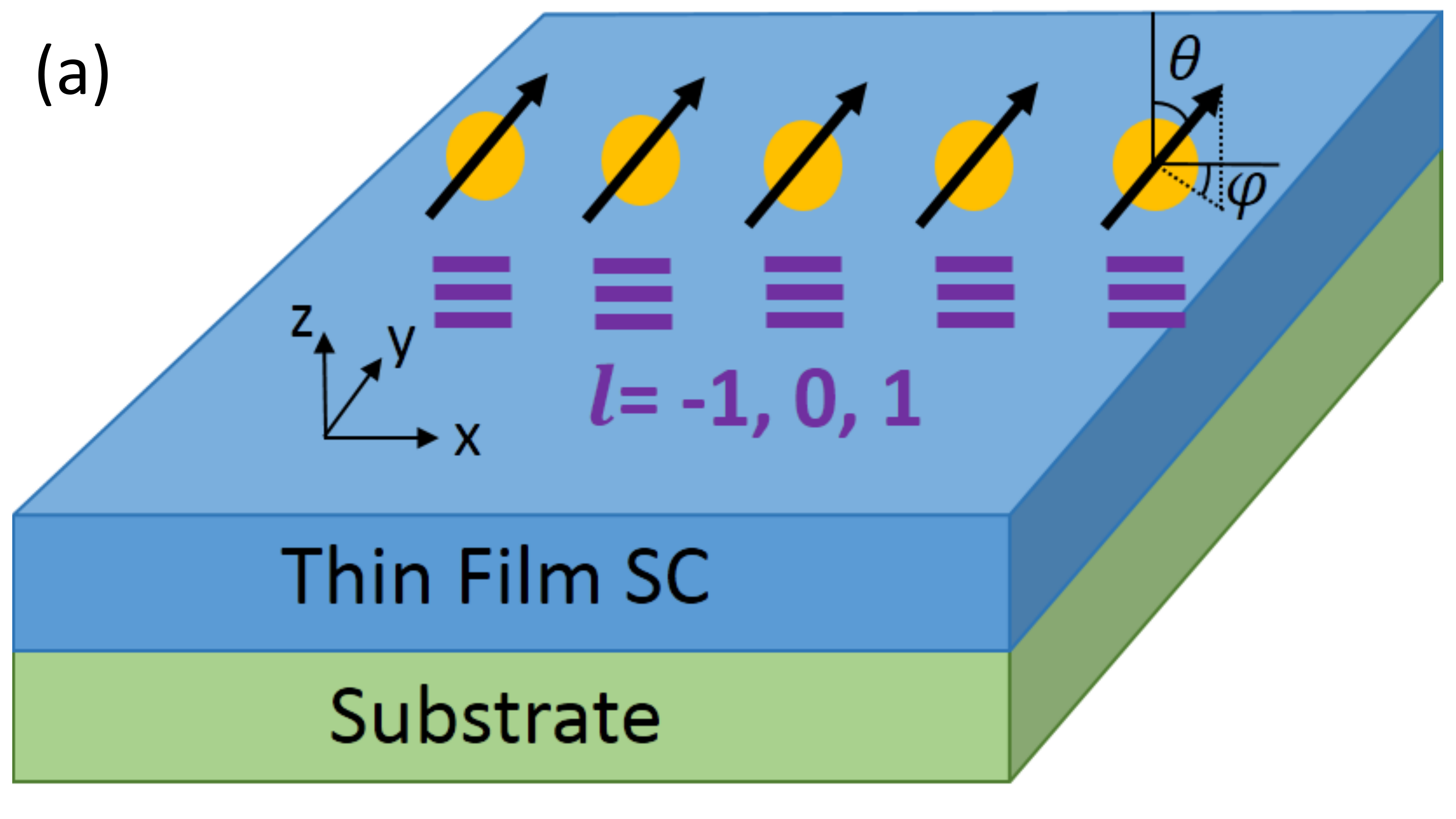}\\
\includegraphics[width=3.3in]{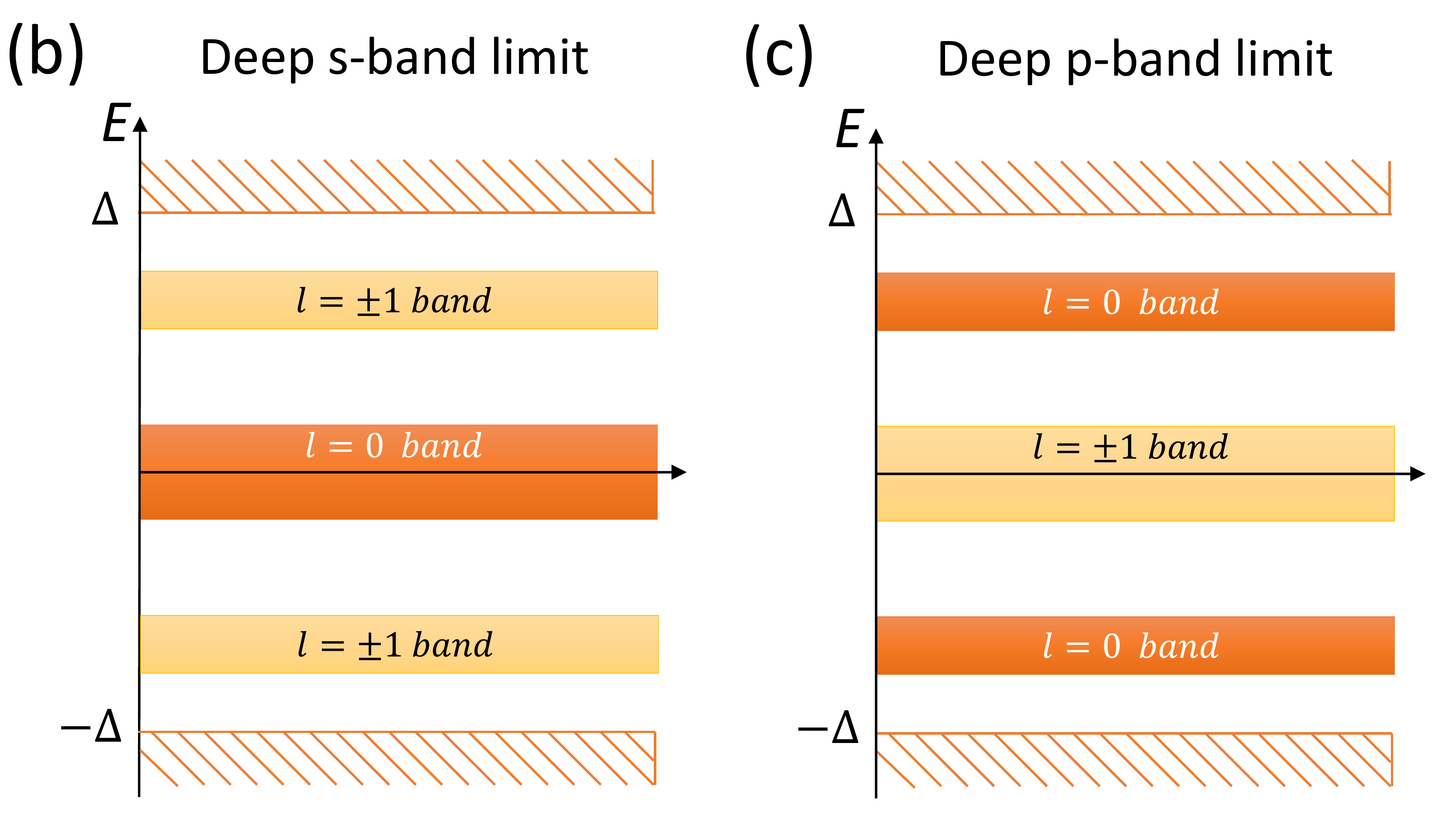}
\caption{(Color online) Schematic setup of the proposed structure supporting multichannel Yu-Shiba-Rusinov chain. Atoms with high magnetic moments form a ferromagnetic chain on the surface of thin film SC~\cite{nadjperge2014, kim2014}. Each magnetic atom creates multiple Yu-Shiba-Rusinov states with distinct angular momentum quantum number $l$. Here we consider $l=-1$, $0$, and $1$ states which forms three bands in a quasi one-dimensional system. In the topological phase, there is an odd number of Majorana zero-energy modes at the opposite ends of the chain. \label{fig: device}}
\end{figure}

In this work we study a realistic model for a chain of magnetic
impurities placed on the surface of an s-wave superconductor, see Fig. \ref{fig: device},
that takes into account multiple scattering channels for the adatoms and the presence of SOC.
Treating the magnetic impurities classically we study the properties
of the bands formed by the hybridization of the
YSR states bound to the different adatoms forming the chain.
The multichannel treatment of the scattering potential of a single impurity
implies that, for the chain, we obtain a multiband model.
In the remainder, to simplify the presentation and to be able to clearly
point out the main qualitative features of the multiband structure
resulting from the hybridization of multichannel YSR states
at different impurity sites, we assume that
the YSR states corresponding
to different values of $|l|$ are well separated in energy.
In this limit we immediately notice a fundamental difference between the
band, $s$-band, formed from the $|l|=0$ states and the bands formed from $|l|>0$ states:
the $s$-band can be assumed to be well separated from all the other bands
and therefore can be treated effectively as a single isolated band;
however, given the degeneracy, in the limit of no SOC, of the $+|l|$, $-|l|$
states, we have that the two bands formed from $+|l|$, and $-|l|$ states are  always very close in energy and
therefore that for bands formed from $|l|>0$ states, a multiband treatment is necessary.
Henceforth, we limit ourselves to the case in which
${\rm max}(|l|)=1$, and consider the minimal model that captures the aforementioned features.

The topological properties of the chain are determined by the band (bands) that is  (are) closest
to the midgap energy of the superconductor. To exemplify the main
properties of the chain of multichannel magnetic impurities we consider two limits:
(i)  the limit in which the $l=0$ YSR states are closest to the midgap energy of the superconductor so that for the chain the
     resulting $s$-band is also the closest to the midgap region,
     see Fig.~\ref{fig: device}~(b); we call this case ``deep~$s$-band'' limit;
(ii) the limit in which the $|l|=1$  YSR states are closest to the midgap energy of the superconductor so that for the chain the
     resulting bands, ``$p$-bands'', are also the closest to the midgap region
     see Fig.~\ref{fig: device}~(c); we call this case ``deep~$p$-band'' limit.
We obtain the topological phase diagram for both the deep~$s$-band and the deep~$p$-band limit.
The presence in the deep~$p$-band limit of two bands close in energy would suggest that
in this limit the topological phase could be strongly suppressed. Contrary to this
na\"ive expectation we find that in the deep~$p$-band limit the
phase space in which the topological phase is present can be even larger than in the deep~$s$-band limit.
We also find that in the multichannel case the
presence of SOC leads to the dependence of the chemical potential on
the direction of the chain magnetization, which, in principle, allows
one to tune between topological and non-topological phases. This is an
important feature for braiding Majoranas~\cite{AliceaBraiding}.

The paper is organized as follows. In Sec.~\ref{sec:2} we introduce
our model and explain the general framework for the calculation. In
Secs.~\ref{sec:s-band} and ~\ref{sec:p-band}, we derive an effective
Hamiltonian and calculate the topological phase diagram as well as
quasi-particle gap for the deep $s$- and $p$-band limits,
respectively. In Sec.\ref{sec:conclusions}  we discuss the qualitative difference between those two
limits and compare their topological phase diagrams. The technical details are presented in the
Appendices.

\section{Theoretical model for multichannel ysr chain}\label{sec:2}

We consider a chain of magnetic impurities separated by a distance
$a$ and placed on top of an $s$-wave superconductor with Rashba
spin-orbit coupling (SOC). The corresponding Hamiltonian describing
an effectively two-dimensional superconducting film with Rashba SOC
reads ($\hbar=1$):
\begin{equation}
H_{\rm SC}=\left[\varepsilon_{\mathbf{k}}\sigma_{0}+\tilde{\alpha}\left(\boldsymbol{\sigma}\times\mathbf{k}\right)\cdot\hat{\mathbf{z}}\right]\tau_{z}+\Delta\sigma_{0}\tau_{x},
\end{equation}
where $\mathbf{k}=(k_{x},\ k_{y})$ is the electron momentum, $\varepsilon_{\mathbf{k}}=k^{2}/2m-\mu$
with $\mu$ the chemical potential and $m$ the effective electron mass, $\tilde{\alpha}$ is the strength
of the Rashba SOC, $\hat{\mathbf{z}}$ is the unit vector normal to
the plane, see Fig.\ref{fig: device}, and $\Delta$ is the superconducting gap. The Pauli matrices
$\sigma_{i}$ and $\tau_{i}$ operate in spin and particle-hole space,
respectively.
The presence of the magnetic impurities is taken into account via the Hamiltonian
\begin{equation}
H_{{\rm imp}}=\sum_{j}V_{j}(\mathbf{r}-\mathbf{R}_{j})\equiv-\sum_{j}\tilde{J}(\mathbf{r}-\mathbf{R}_{j})\left(\mathbf{S}_{j}\cdot\boldsymbol{\sigma}\right)\tau_{0}
\end{equation}
where $V_{j}$ are the individual impurity magnetic potentials with
$\mathbf{R}_{j}$, $\mathbf{S}_{j}$ and $\tilde{J}$ being the impurity
position, its classical spin and its exchange coupling to the host quasiparticles,
respectively. In order to find the band structure of the chain of YSR states induced by the magnetic impurities,
we need to solve the Schrodinger equation:
\begin{equation}
\left[H_{{\rm SC}}+H_{{\rm imp}}(\mathbf{r})\right]\psi(\mathbf{r})=E\psi(\mathbf{r}).\label{eq:Schrodinger}
\end{equation}
Here $\psi(\mathbf{r})$ is the Nambu spinor $\left(\begin{array}{cccc}
\psi_{\uparrow}(\mathbf{r}), & \psi_{\downarrow}(\mathbf{r}), & \psi_{\downarrow}^{\dagger}(\mathbf{r}), & -\psi_{\uparrow}^{\dagger}(\mathbf{r})\end{array}\right)^{T}$. The spectrum of the subgap states is determined by the pole of the $T$-matrix~\cite{balatsky:373}. In terms of the Green's function for the superconductor $G=\left[E-H_{{\rm SC}}\right]^{-1}$,
the eigenvalue problem for the subgap states is given by $(1-GH_{imp})\psi(\mathbf{r})=0$.
We note that the Green's function of an s-wave superconductor in the
presence of SOC has both even- and odd-parity components \cite{Gorkov01},
and can be written as $G(\mathbf{k};E)=\sum_{l}G_{l}(k;E)e^{il\theta_{\mathbf{k}}}$
where $k=|\mathbf{k}|$ and $\theta_{\mathbf{k}}=\arctan k_{x}/k_{y}$.
In the presence of Rashba SOC, the ${G_l}$ are non-zero only for $l=-1,0,1$
and are given by:
\begin{align}
G_{-1}(k;E) & =\frac{1}{2}\sum_{\lambda=\pm}\left(-i\lambda\right)\frac{E\sigma_{+}\tau_{0}+\Delta\sigma_{+}\tau_{x}+\varepsilon_{\lambda}(k)\sigma_{+}\tau_{z}}{E^{2}-\varepsilon_{\lambda}^{2}(k)-\Delta^{2}}\label{eq:gf1}\\
G_{0}(k;E) & =\frac{1}{2}\sum_{\lambda=\pm}\frac{E\sigma_{0}\tau_{0}+\Delta\sigma_{0}\tau_{x}+\varepsilon_{\lambda}(k)\sigma_{0}\tau_{z}}{E^{2}-\varepsilon_{\lambda}^{2}(k)-\Delta^{2}}\label{eq:gf2}\\
G_{1}(k;E) & =\frac{1}{2}\sum_{\lambda=\pm}\left(i\lambda\right)\frac{E\sigma_{-}\tau_{0}+\Delta\sigma_{-}\tau_{x}+\varepsilon_{\lambda}(k)\sigma_{-}\tau_{z}}{E^{2}-\varepsilon_{\lambda}^{2}(k)-\Delta^{2}}\label{eq:gf3}
\end{align}
where $\sigma_{\pm}=\left(\sigma_{x}\pm i\sigma_{y}\right)/2$, and
$\varepsilon_{\lambda}(k)=k^{2}/2m-\lambda\tilde{\alpha}k-\mu$ are the dispersions
of the helical bands with $\lambda=\pm$.

Transforming to momentum space, Eq.~(\ref{eq:Schrodinger}) can then be rewritten as:
\begin{equation}
\psi_{i}(\mathbf{k})=\sum_{j}\left[e^{i\mathbf{k}\cdot(\mathbf{R}_{i}-\mathbf{R}_{j})}G(\mathbf{k};E)\int\frac{d\mathbf{p}}{(2\pi)^{2}}V_{j}(\mathbf{k-p})\psi_{j}(\mathbf{p})\right]\label{eq:chain_Eq}
\end{equation}
where $\psi_{i}(\mathbf{k})$ is the Fourier transform of the wave
function centered at $\mathbf{R}_{i}$. Assuming that at the Fermi
surface the scattering potential is weakly dependent on $p\equiv |\mathbf{p}|$ we have $V_{j}(\mathbf{k-p})\approx V_{j}(\theta_{\mathbf{k}}-\theta_{\mathbf{p}})$, where
\begin{equation}
V_{j}(\theta_{\mathbf{k}}-\theta_{\mathbf{p}})=\left(\begin{array}{cc}
-\tilde{J}(\theta_{\mathbf{k}}-\theta_{\mathbf{p}})\mathbf{S}_{j}\cdot\boldsymbol{\sigma} & 0\\
0 & -\tilde{J}(\theta_{\mathbf{p}}-\theta_{\mathbf{k}})\mathbf{S}_{j}\cdot\boldsymbol{\sigma}
\end{array}\right).
\end{equation}
The magnetic potential can be decomposed into independent angular
momentum channels: $V_{j}(\theta_{\mathbf{k}}-\theta_{\mathbf{p}})=\sum_{l}V_{j,l}e^{il(\theta_{\mathbf{k}}-\theta_{\mathbf{p}})}$,
so that
\begin{equation}
V_{j,l}=\left(\begin{array}{cc}
-\tilde{J}_{l}\mathbf{S}_{j}\cdot\boldsymbol{\sigma} & 0\\
0 & -\tilde{J}_{-l}\mathbf{S}_{j}\cdot\boldsymbol{\sigma}
\end{array}\right).
\end{equation}
with $\tilde{J}_l$ the angular momentum components of $\tilde{J}(\theta_{\mathbf{k}}-\theta_{\mathbf{p}})$.
Since $V_{j}(\theta_{\mathbf{k}}-\theta_{\mathbf{p}})$ is Hermitian
and an even function of $\theta_{\mathbf{k}}-\theta_{\mathbf{p}}$, we
have $\tilde{J}_{l}=\tilde{J}_{-l}$ and $V_{j,l}=-\tilde{J}_{l}\left(\mathbf{S}_{j}\cdot\boldsymbol{\sigma}\right)\tau_{0}$.
We assume that $l=0$, the $s$ channel, and $l=\pm1$, the $p$ channels,
are the dominant scattering channels: $V_{j}=\sum_{l=0,\pm1}V_{j,l}$.
For the impurity chain arranged along the $\hat{\mathbf{x}}$ direction in a ferromagnetic pattern,
i.e., $\mathbf{S}_{j}=(\cos (\varphi)\sin \theta, \sin (\varphi)\sin \theta, \cos(\theta))$, (illustrated in Fig. \ref{fig: device}(a)),
Eq.~(\ref{eq:chain_Eq}) in the angular mode representation takes the form
\begin{equation}
\overline{\psi}_{i,l}=\sum_{j,l'}\mathcal{G}_{l-l'}^{ij}(E)V_{l'}\overline{\psi}_{j,l'}\label{eq: multichannel_Eq}
\end{equation}
where an overline indicates an integration over momentum amplitude:
$\overline{\psi}_{i,l}\equiv\int\frac{kdk}{2\pi}\psi_{i,l}(k)$, with $\psi_{i,l}$ the angular momentum components of $\psi_{i}(\mathbf{k})$
($\psi_{i}(\mathbf{k})=\sum_{l}\psi_{i,l}(k)e^{il\theta_{\mathbf{k}}}$), and
$\mathcal{G}_{l-l'}^{ij}(E)\equiv\int\frac{d\mathbf{k}}{(2\pi)^{2}}e^{ikx_{ij}\cos\theta_{\mathbf{k}}}e^{-i(l-l')\theta_{\mathbf{k}}}G(\mathbf{k};E)$
with $x_{ij}\equiv x_{i}-x_{j}$. Using Eq.~(\ref{eq:gf1})-(\ref{eq:gf3})
one can rewrite $\mathcal{G}_{l-l'}^{ij}(E)$ in terms of the dimensionless integrals:
\begin{equation}
I_{n,\lambda}(x;E)=\frac{N_{\lambda}}{2\pi^{2}N_{F}}\int_{-\pi}^{\pi}d\theta_{\mathbf{k}}\int_{-D}^{D}d\varepsilon\frac{e^{ik_{\lambda}(\varepsilon)x\cos\theta_{\mathbf{k}}}e^{in\theta_{\mathbf{k}}}\Delta }{E^{2}-\varepsilon^{2}-\Delta^{2}},\label{eq:if1}
\end{equation}
\begin{equation}
K_{n,\lambda}(x;E)=\frac{N_{\lambda}}{2\pi^{2}N_{F}}\int_{-\pi}^{\pi}d\theta_{\mathbf{k}}\int_{-D}^{D}d\varepsilon\frac{e^{ik_{\lambda}(\varepsilon)x\cos\theta_{\mathbf{k}}}e^{in\theta_{\mathbf{k}}}\varepsilon}{E^{2}-\varepsilon^{2}-\Delta^{2}},\label{eq:if2}
\end{equation}
where $D$ is an energy cut-off, $k_{\lambda}(\varepsilon)=k_{F,\lambda}+\varepsilon/v_{F,\lambda}$
with $k_{F,\lambda}=k_{F}\left(\sqrt{1+\alpha^{2}}+\lambda\alpha\right)$, $v_{F,\lambda}=v_{F}\sqrt{1+\alpha^{2}}$,
$k_{F}=\sqrt{2m\mu}$, $v_{F}=k_{F}/m$. $\alpha\equiv\tilde{\alpha}m/k_{F}$ is
the dimensionless SOC coupling. $N_{\lambda}=\frac{m}{2\pi}\left[1+\lambda\frac{\alpha}{\sqrt{1+\alpha^{2}}}\right]$
is the density of states of the $\lambda$ helical band at the Fermi level
in the normal state, and $N_{F}=(N_{+}+N_{-})/2$. The analytic results for the above integrals
in the limit $D\rightarrow\infty$ and for $\mathcal{G}_{l-l'}^{ij}(E)$ are presented
in Appendix~\ref{app:A} and \ref{app:B}. For convenience, we define the dimensionless
exchange couplings $J_{l}\equiv\tilde{J}_{l}|\mathbf{S}|\pi N_{F}$ which will be used henceforth.

It is convenient to rewrite Eq.~(\ref{eq: multichannel_Eq}) in the following
form:
\begin{equation}
\sum_{j}\mathbf{M}^{ij}(E)\overline{\Psi}_{j}=0\label{eq:mp}
\end{equation}
where $\overline{\Psi}_{j}=(\overline{\psi}_{i,-1},\,\overline{\psi}_{i,0},\,\overline{\psi}_{i,1})^{T}$
is a 12 dimensional spinor, and the matrix $\mathbf{M}^{ij}(E)$ is defined as   $\mathbf{M}_{l,l'}^{ij}=\delta_{i,j}\delta_{l,l'}-\mathcal{G}_{l-l'}^{ij}(E)V_{l'}$. Here the local part of the matrix $\mathbf{M}_{l,l'}^{ii}$ determines the YSR spectrum of a single magnetic atom~\cite{kim2014b} whereas the non-local part $\mathbf{M}_{l,l'}^{ij}$ describes the hybridization between YSR states induced by the magnetic atoms at $i$ and $j$ sites. For an equally spaced magnetic atom chain with distance $a$ between the two nearest atoms, this hybridization leads to the formation of the YSR bands. In the limit of $k_{F}a\gg1$, which we consider henceforth, the hopping energy scale is proportional to $1/\sqrt{k_{F}a}$ and, thus, the bandwidth $W$ is small, i.e. $W\ll\Delta$. In this limit, the bands maintain the character of the single impurity YSR states and, thus, we refer to them as $s$ or $p$-bands. Strictly speaking, SOC mixes different angular momentum states but, since we assume that $\alpha\ll1$, this terminology is justified.

When $s$ and $p$ bands are well-separated by a gap that is much larger than the temperature, see Fig.~\ref{fig: device}(b) and (c), the problem can be considerably simplified by integrating out the higher-energy bands (i.e. by taking into account virtual scattering processes to the bands higher in energy). In the following, we consider two limiting cases corresponding to the deep $s$- and $p$-band limits and discuss the corresponding topological phase diagrams. We show that these two cases are qualitatively different since deep $p$ band limit consists of two bands originating from the $l=\pm 1$ YSR states.

\section{deep $S$ band limit}\label{sec:s-band}

\subsection{Derivation of the Effective Hamiltonian}

We first consider the deep $s$-band limit such that the energy of the $l=0$ state is close to the midgap,
i.e., $J_{0}\sim1$ with the on-site energy $\epsilon_{0}\approx\Delta(1-J_{0})+\mathcal{O}(\alpha^2) \rightarrow 0$.
Provided $J_{1}\ll J_{0}$ the $l=0$ band is well separated from the $p$ bands, i.e., the bandwidth $W\ll\Delta\left|J_{0}-J_{1}\right|$.
After integrating out the $l=\pm1$ states, we obtain a
tight-binding description for the single $s$-band with the $p$ channels
taken into account perturbatively, by allowing for virtual transitions through the $p$ channels. This can be done by first solving for
 $\overline{\psi}_{i,\pm1}$ using Eq.~(\ref{eq:mp})
\begin{align}
\overline{\psi}_{i,-1}&=-(\mathbf{M}_{-1,-1}^{ii})^{-1}(\mathbf{M}_{-1,0}^{ii}\overline{\psi}_{i,0}+\sum_{j\neq i,l}\mathbf{M}_{-1,l}^{ij}\overline{\psi}_{j,l}),\nonumber\\
\overline{\psi}_{i,1}&=-(\mathbf{M}_{1,1}^{ii})^{-1}(\mathbf{M}_{1,0}^{ii}\overline{\psi}_{i,0}+\sum_{j\neq i,l}\mathbf{M}_{1,l}^{ij}\overline{\psi}_{j,l}),\label{eq:s2}
\end{align}
and substituting above expressions into the equation for $l=0$ component and keeping terms up to the linear order in inter-site coupling, we obtain
\begin{equation}
\sum_{j}\mathbf{M}_{s}^{ij}(E)\overline{\psi}_{j,0}=0,\label{eq: s_channel_chain_Eq0}
\end{equation}
where the matrix $\mathbf{M}_{s}^{ij}(E)$ is given in the
Appendix \ref{app:C}. In order to solve Eq.~\eqref{eq: s_channel_chain_Eq0} analytically, we expand the local on-site matrix to the linear order in $E$ around $E=0$, assuming that $\epsilon_0\rightarrow 0$,
$$\mathbf{M}_{s}^{ii}(E)\approx\mathbf{M}_{s}^{ii(0)}-\mathbf{M}_{s}^{ii(1)}\cdot E,$$ and set $E=0$ in the inter-site matrix: $$\lim_{E\rightarrow0}\mathbf{M}_{s}^{i\neq j}(E)\equiv\mathbf{M}_{s}^{i\neq j}(0).$$ In doing so we ignore terms $\mathcal{O}\left(1/k_{F}a\right)\ll 1$ and $\mathcal{O}\left( \frac{E}{\Delta\sqrt{k_{F}a}}\right).$ With these approximations,
Eq.(\ref{eq: s_channel_chain_Eq0}) can be written as
\begin{equation}
\sum_{j}H_{s}^{ij}\overline{\psi}_{j}=E\overline{\psi}_{i},\label{eq: s_band_linear_chain_Eq}
\end{equation}
where the local and non-local contributions are given by $H_{s}^{ii}=\left(\mathbf{M}_{s}^{ii(1)}\right)^{-1}\mathbf{M}_{s}^{ii(0)}$
and $H_{s}^{ij}=\left(\mathbf{M}_{s}^{ii(1)}\right)^{-1}\mathbf{M}_{s}^{i\neq j}(0)$, respectively.
The tight-binding Hamiltonian $\mathcal{H}_{s}(i,j)$
is obtained by projecting Eq. (\ref{eq: s_band_linear_chain_Eq})
onto the local YSR states: $\left(\begin{array}{cc}
\varphi_{+}, & \varphi_{-}\end{array}\right)^{T}$ where $\varphi_{\pm}$ are the eigen spinors of the single-impurity
bound states with energy $\pm\epsilon_0$. The local basis can be found
by solving the single-site equation $\mathbf{M}_{s}^{ii}(E)\varphi_{\pm}=0$
as a special case of Eq. (\ref{eq: s_channel_chain_Eq0}), where the
bound state energies are determined from $\mathrm{Det}\left[\mathbf{M}_{s}^{ii}(E)\right]=0$, see Appendix~\ref{app:D}.

We first assume that the magnetic atom moments are aligned ferromagnetically along
a) $\hat{\mathbf{z}}$- (out-of-plane), b) $\hat{\mathbf{x}}$- (along the chain direction), and c) $\hat{\mathbf{y}}$- (in-plane but normal to the chain) axis, and present explicit expressions for the corresponding effective Hamiltonian. We note that due to the presence of SOC, the effective Hamiltonian is anisotropic which can be readily seen already at the single-impurity level~\cite{kim2014b}.

In the a) and b) cases, by transforming the effective tight-binding
Hamiltonian $\mathcal{H}_{s}(i,j)$ to momentum space, we find that the corresponding Bogoliubov-de Gennes (BdG) Hamiltonian reads
\begin{equation}
\frac{\mathcal{H}_{s}^{\hat{z}(\text{or}\,\hat{x})}(k)}{\Delta}=\left(\begin{array}{cc}
h_{z(x)}(k) & \tilde{\Delta}_{z(x)}(k)\\
\tilde{\Delta}_{z(x)}^{*}(k) & -h_{z(x)}(k)
\end{array}\right).\label{eq: s_band_H}
\end{equation}
To order $\alpha^{2}$ and $\alpha/\sqrt{k_{F}a}$, the effective
hopping energy is given by
\begin{align}\label{eq:l=0 h}
h_{z}( & k)  \simeq \epsilon_{z}+\frac{1}{2}\left[I_{0,+}(k)+I_{0,-}(k)\right] \\
& \stackrel{k \rightarrow 0}{\approx} h_z^{(0)} + h_z^{(2)}k^2.\nonumber\\
h_{x}( & k)\simeq\epsilon_{x}+\frac{1}{2}\left[I_{0,+}(k)+I_{0,-}(k)\right]\\
& \stackrel{k \rightarrow 0}{\approx} h_x^{(0)} + h_x^{(2)}k^2,\nonumber
\end{align}
where the functions $I_{0,\pm}(k)$ and $I_{2,\pm}(k)$ are obtained by taking the $E\rightarrow 0$ limit in Eqs.\eqref{eq:I0}. The on-site energy is
\begin{align}\label{eq: l=0 chem}
\epsilon_{z} & \simeq  \frac{1-J_{0}}{J_0}+\frac{\alpha^{2}J_{1}(2-J_{0}+J_{1})}{J_{0}(1+J_{1})^{2}},\nonumber \\ \,\\
\epsilon_{x} & \simeq  \frac{1-J_{0}}{J_0}-\frac{\alpha^{2}J_{1}^{2}\left[2(1-J_{0})+(1-J_{1}^{2})\right]}{J_{0}(1-J_{1}^{2})^{2}}\nonumber
\end{align}
The effective p-wave pairing take the form
\begin{align}\label{eq:l=0 gap}
\tilde{\Delta}_{z}(k)\simeq\tilde{\Delta}_{x}(k)\simeq & \frac{i}{2}\left[K_{1,+}(k)-K_{1,-}(k)\right]\nonumber \\
- & \frac{i\alpha J_{1}}{1+J_{1}}\left[K_{1,+}(k)+K_{1,-}(k)\right] \\
\stackrel{k \rightarrow 0}{\approx} & \Delta^{(1)}k.
\end{align}
The functions $I_{n,\pm}(k)\equiv I_{n,\pm}(k,E=0)$ and $K_{n,\pm}(k)\equiv K_{n,\pm}(k,E=0)$ are defined in the Appendix~\ref{app:A}, see Eqs.(\ref{eq:I0}-\ref{eq:K1}). The functions $h(k)$ and $\tilde{\Delta}(k)$ have the following properties $h(k)=h(-k)$ and $\tilde{\Delta}(-k)=-\tilde{\Delta}(k)$, and, thus, the gap is generically vanishing at $k=0, \pi/a$. Moreover, the effective pairing $\tilde{\Delta}(k)$ is vanishing for $\alpha \rightarrow 0$
analogous to the semiconductor nanowire proposal~\cite{Lutchyn2010, Oreg2010} in that the SOC controls the excitation gap.

To have a better understanding of the Hamiltonian structure in Eq.~\eqref{eq: s_band_H}, it is instructive to perform a perturbative expansion of $h(k)$ and $\tilde{\Delta}(k)$, for example, around $k=0$, see Eqs. \eqref{eq:l=0 h} and \eqref{eq:l=0 gap} where the expressions for $h_{z,x}^{(0,2)}$ and $\Delta^{(1)}$ are given in the Appendix~\ref{app:G}. In the limit $a/\xi_{0} \rightarrow 0$, where $\xi_{0}=v_{F}/\Delta$ is the superconducting coherence length, these functions have singular points for some values of $k_F a$ which is a consequence of the long-range nature of the hopping matrix element in the effective Hamiltonian. The presence of a finite coherence length $\xi_{0}$, however, regularizes the singularities. Nevertheless, such a strong dependence on $k_Fa$ leads to significant variations of the effective mass and Fermi velocity. Another important feature that we find is that the effective Hamiltonian is anisotropic due to the SOC (cf. Eq.\eqref{eq: l=0 chem}) which might be helpful to drive the topological transition by changing the direction of the magnetization of the impurities forming the chain. We note that this effect is absent for $J_1=0$, in which case we recover the results of Ref.~\cite{brydon2014}. Thus, the dependence of the effective chemical potential on the angle $\theta$, which is the only tuning parameter in the Hamiltonian~\eqref{eq: s_band_H}, is a feature of the multichannel magnetic impurity model.

Finally, we discuss the case in which $\mathbf{S}\parallel\hat{\mathbf{y}}$. In this case the magnetic exchange energy term and the SOC term commute, and, as a result, the system is gapless. Thus, this case is not interesting from the point of view of Majorana physics.

\subsection{Topological Properties}

\begin{figure}[htb]
\includegraphics[height=1.79in]{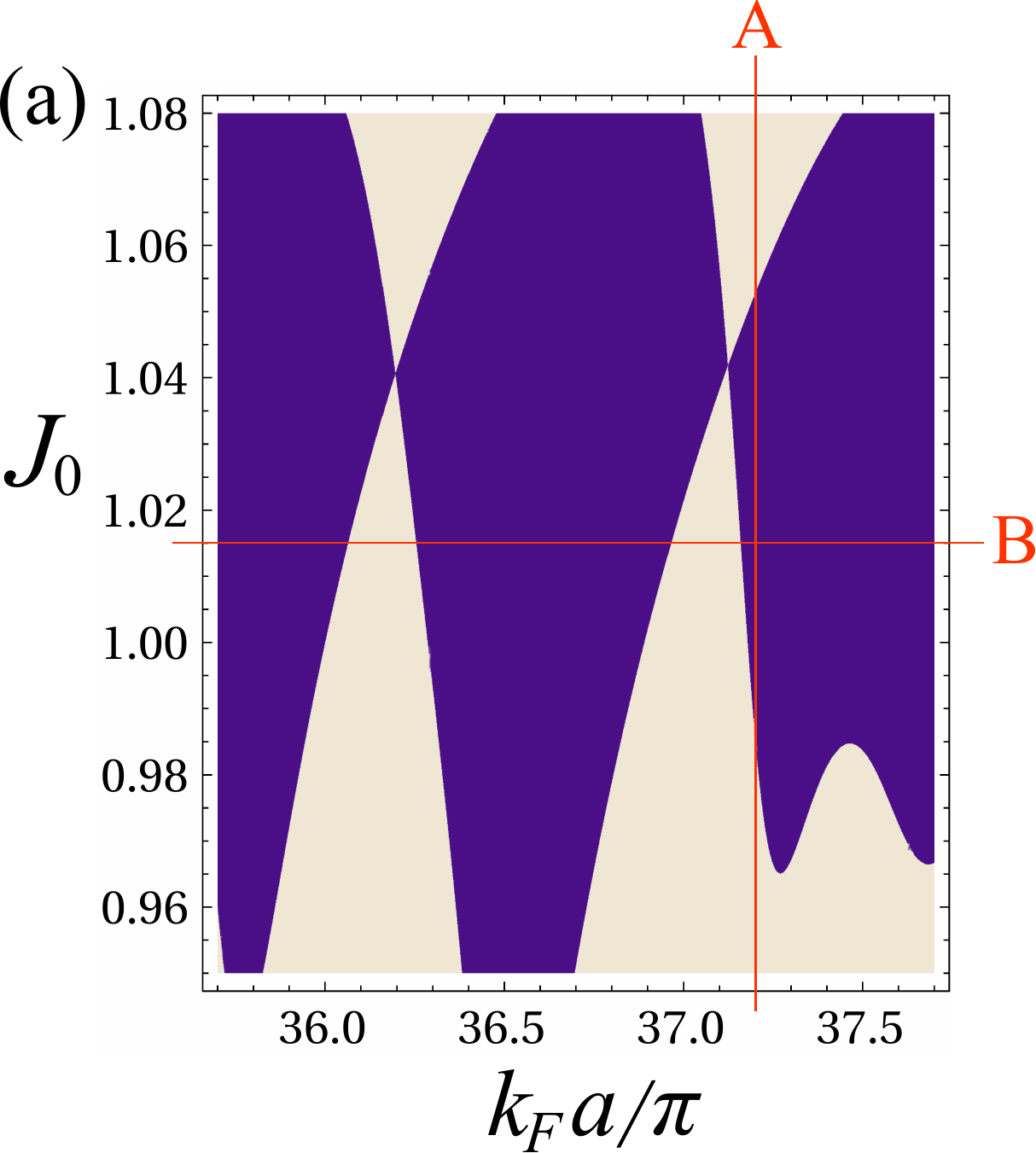}\hfill{}\includegraphics[height=1.79in]{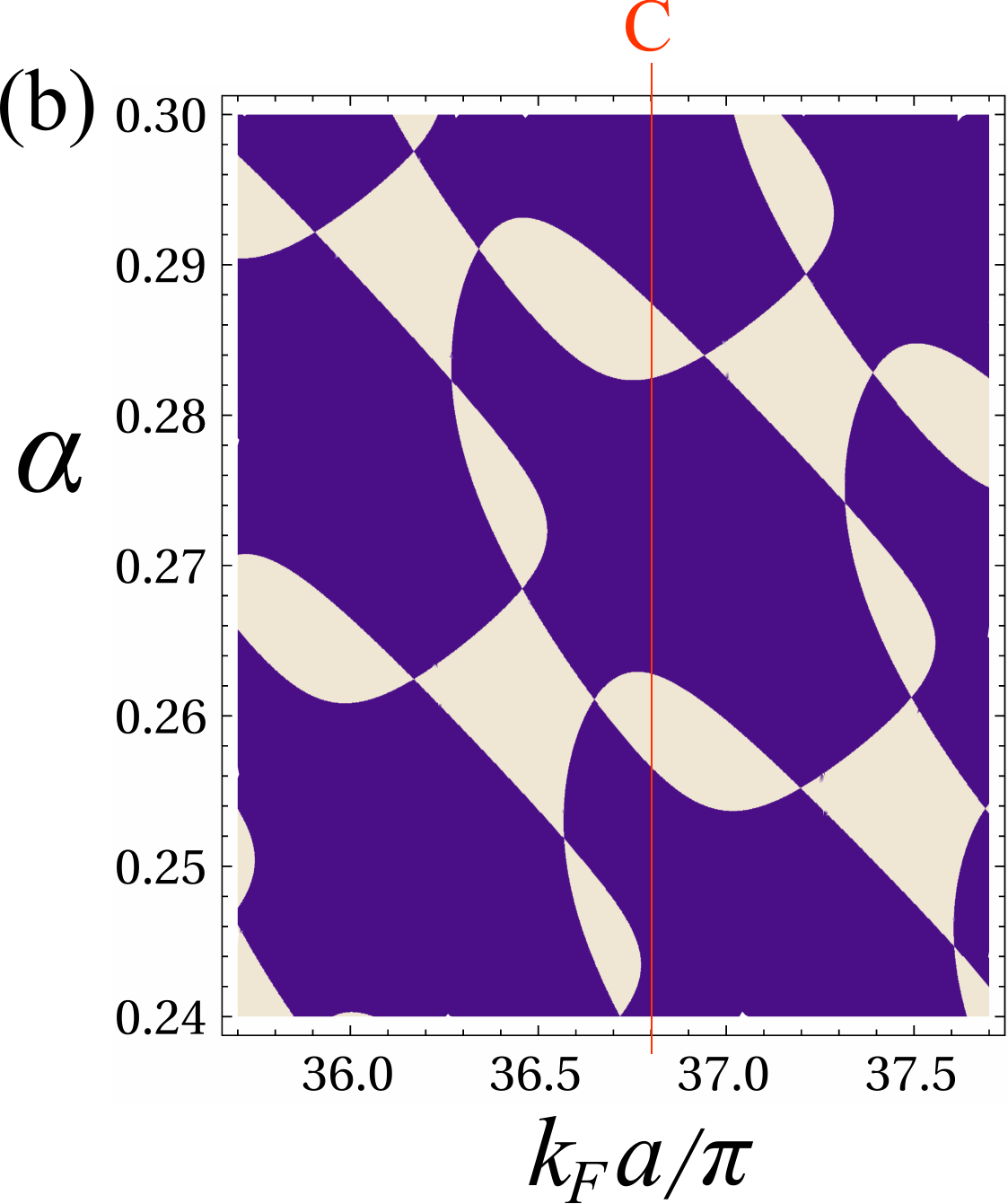}\hfill{}

\includegraphics[height=1.62in]{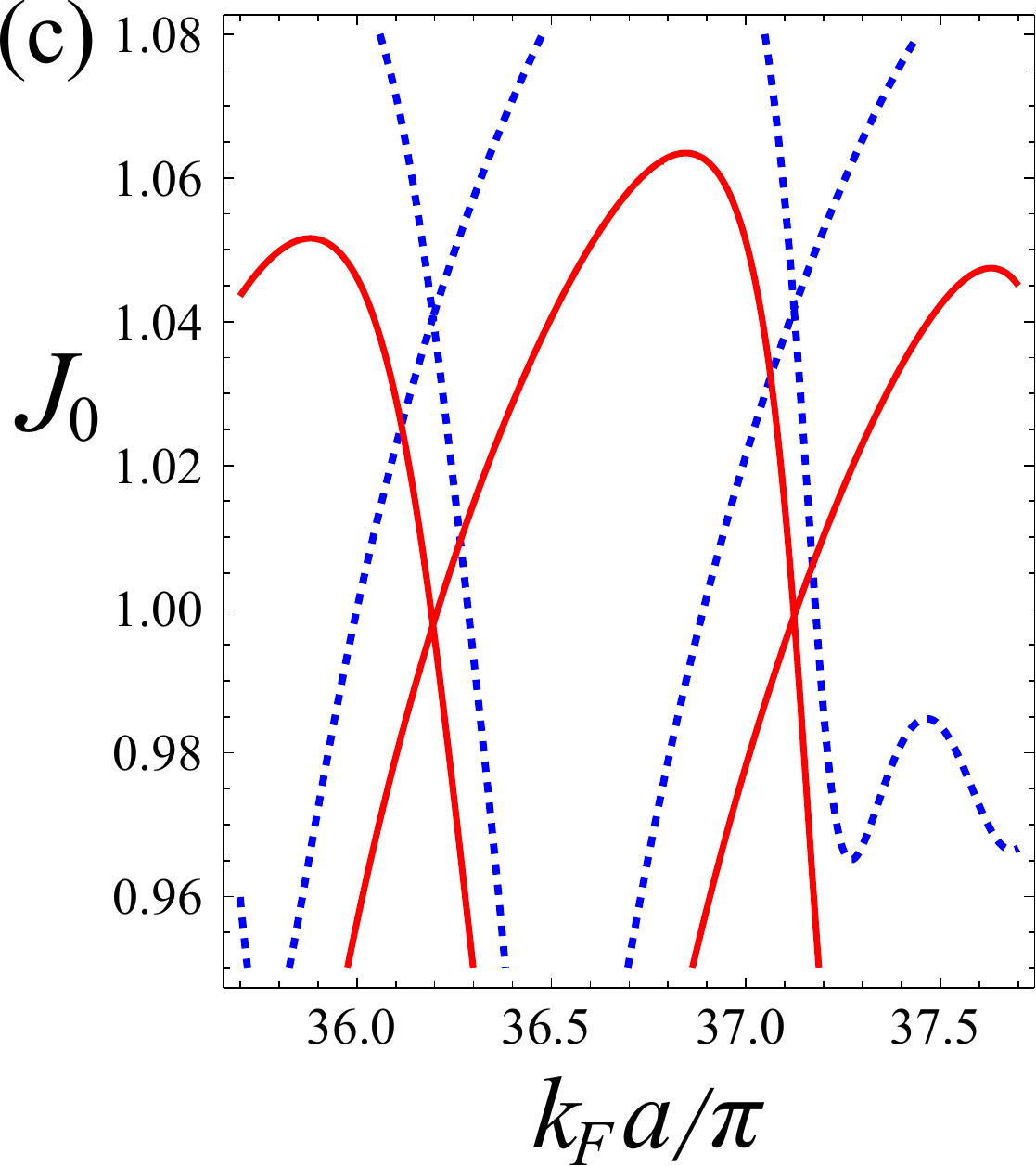}\hfill{} \hfill{}\includegraphics[height=1.65in]{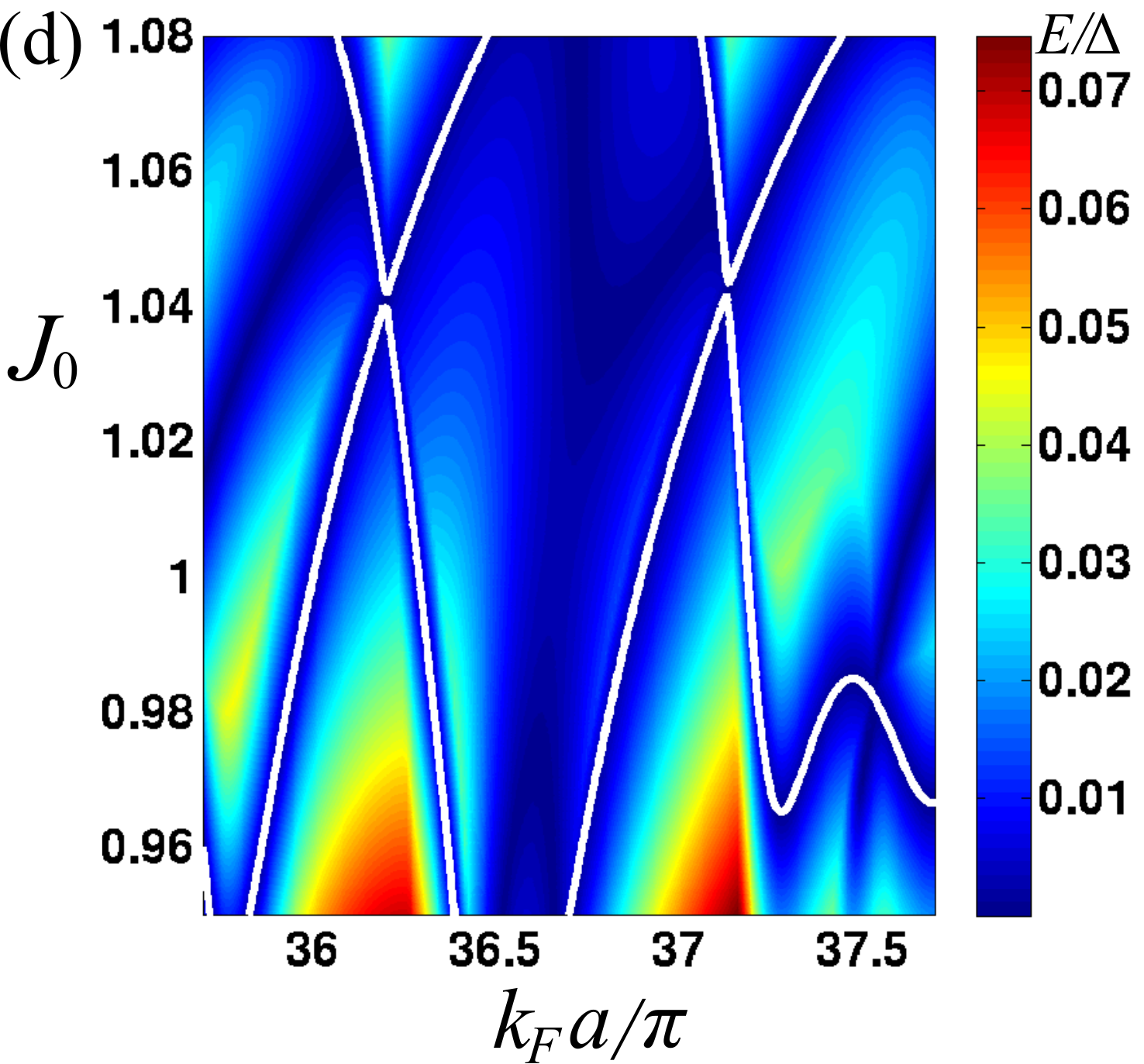}
\caption{(Color online) Topological phase diagram for the deep $s$ band as a function of physical parameters. The dark and light colors represent topologically nontrivial and trivial phases, respectively. (a) Topological phase diagram for the magnetization in $\hat{\mathbf{z}}$ direction, as a function of
$J_{0}$ and $k_{F}a$ for $\alpha=0.3$, $J_{1}=0.4$, and $\xi_0=2a$.
(b) Topological phase diagram for the magnetization in $\hat{\mathbf{z}}$ direction as a function of $\alpha$ and $k_{F}a$
for $J_{0}=1.025$, $J_{1}=0.4$, and $\xi_0=2a$.
(c) The phase boundary for the magnetization in $\hat{\mathbf{z}}$ direction (blue dashed line) and for the magnetization in $\hat{\mathbf{x}}$ direction (red solid line) indicates that by changing the magnetization one can drive the topological phase transition.
(d) Calculated quasiparticle excitation gap for the parameter regime in the phase diagram (a) with the phase boundary indicated by white line. \label{fig: Fig2}}
\end{figure}

Having derived the effective Hamiltonian (\ref{eq: s_band_H}), we can now calculate the topological phase diagram. The Hamiltonian \eqref{eq: s_band_H} for a generic direction of magnetization is in the symmetry class D~\cite{altland1997, schnyder2008, kitaev2008}, and, thus, is characterized by the $Z_2$ topological invariant, the so-called Majorana number $\mathcal{M}$ \cite{Kitaev2001}:
\begin{align}
\mathcal{M}=\mathrm{sgn}\left[h(0)h(\pi/a)\right].
\end{align}
The system is in the topological superconducting phase when $\mathcal{M}=-1$, whereas
$\mathcal{M}=+1$ indicates a non-topological phase. We obtain the topological phase diagram by calculating $\mathcal{M}$.

Figure~\ref{fig: Fig2}~(a) shows the topological phase diagram in the $(k_Fa, J_0)$ plane for the deep~$s$-band limit
for the case in which the magnetic moments of the impurities forming the chain are aligned
along the $z$ direction and $\alpha=0.3$. The range of values of $k_Fa$ has been chosen
so that the inequality $1/\sqrt{k_Fa}\ll 1$, on which the treatment of the previous section relies,
is well satisfied. From Fig.~\ref{fig: Fig2}~(a) we see that, for $\alpha=0.3$ there is a large
fraction of the $(k_Fa, J_0)$ in which the chain is expected to be in a topological phase characterized
by odd number of Majoranas at its ends.
It is interesting to ask how the topological phase diagram is affected by the strength of the SOC.
This question is addressed by the results presented in  Fig.~\ref{fig: Fig2}~(b) that shows
the dependence of ${\cal M}$ on $\alpha$ and $k_Fa$ for a fixed value of $J_0$.
Again we notice that there is a large fraction of the $(\alpha, k_Fa)$ space in which the chain
is expected to be in a topological phase.

Experimentally it can be challenging to vary  in a controlled way parameters such as $\alpha$, $J_0$, and $k_Fa$
and therefore to verify the theoretical predictions shown in Fig.~\ref{fig: Fig2}~(a),~(b).
However, our multichannel treatment, contrary to the single channel treatment
\cite{pientka2013, brydon2014, Heimes15},
reveals that,  the topological character of the chain state also depends on the direction
of the magnetization. This is shown in  Fig.~\ref{fig: Fig2}~(c) in which we can observe
that the boundaries of the topological phase in the $(k_Fa, J_0)$ plane
are different depending on the direction, $z$ or $x$, of the magnetic moment of the impurities
forming the chain. This result can be easily understood by considering that the on-site energy,
Eqs.\eqref{eq: l=0 chem}, depends on the direction of the impurity magnetization.
The dependence of the topological index on the direction of the chain magnetization
is very important because it allows, in principle, to tune the chain in and out
of a topologically nontrivial phase by varying a quantity that can be tuned
and controlled experimentally by using an external magnetic field.
The appearance, disappearance, of a zero bias peak as a function of the direction of
the magnetization of the chain according to theoretical predictions like the ones
presented in Fig.~\ref{fig: Fig2}~(c) would provide
compelling evidence of the Majorana character of the observed zero energy states.

\begin{figure*}[h!!!!t!!!!b]
\includegraphics[width=2.1in]{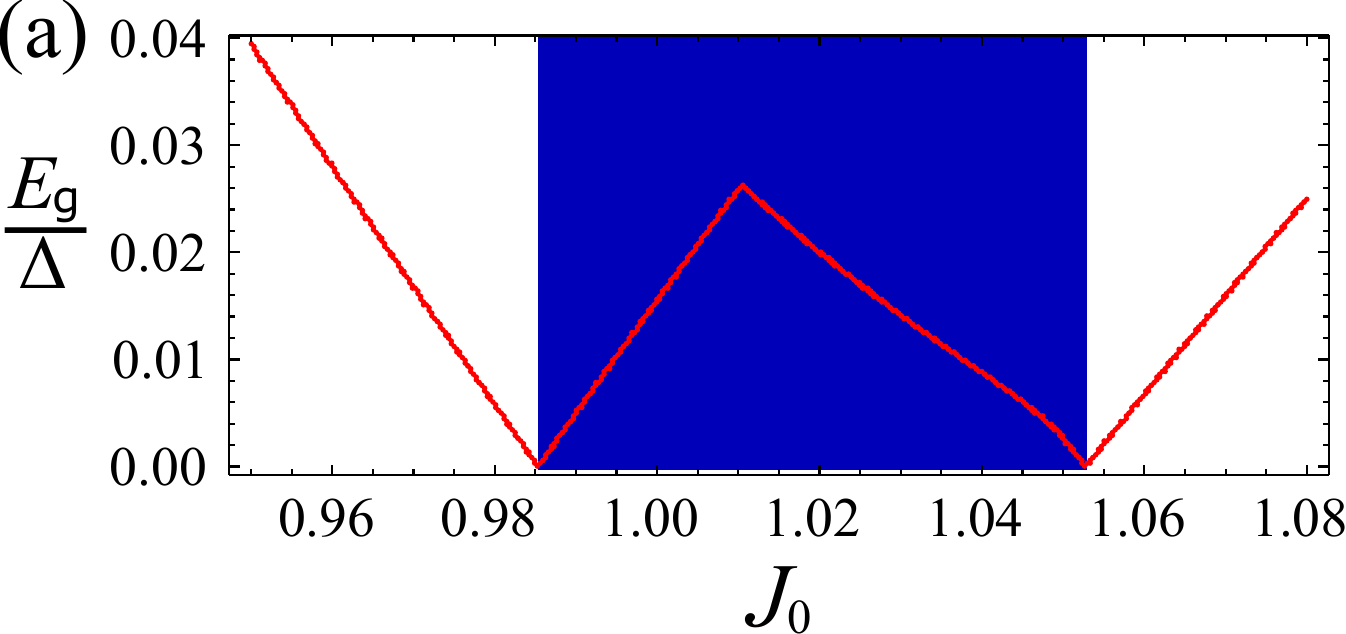}\hfill{} \includegraphics[width=2.15in]{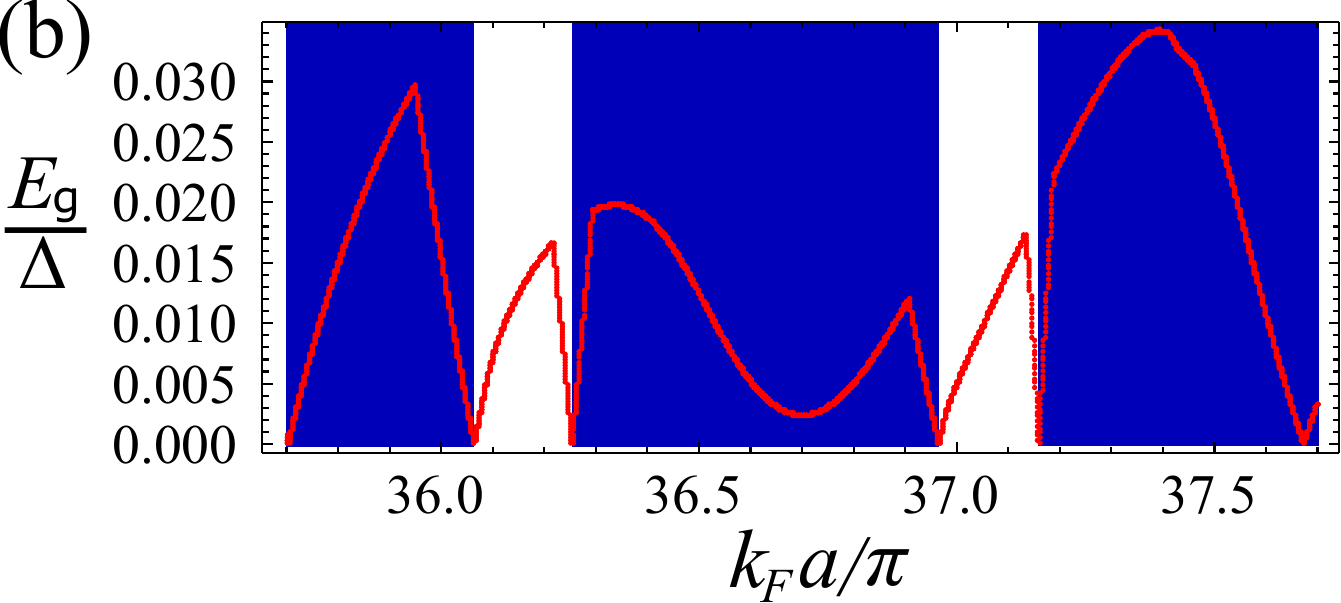}\hfill{} \includegraphics[width=2.15in]{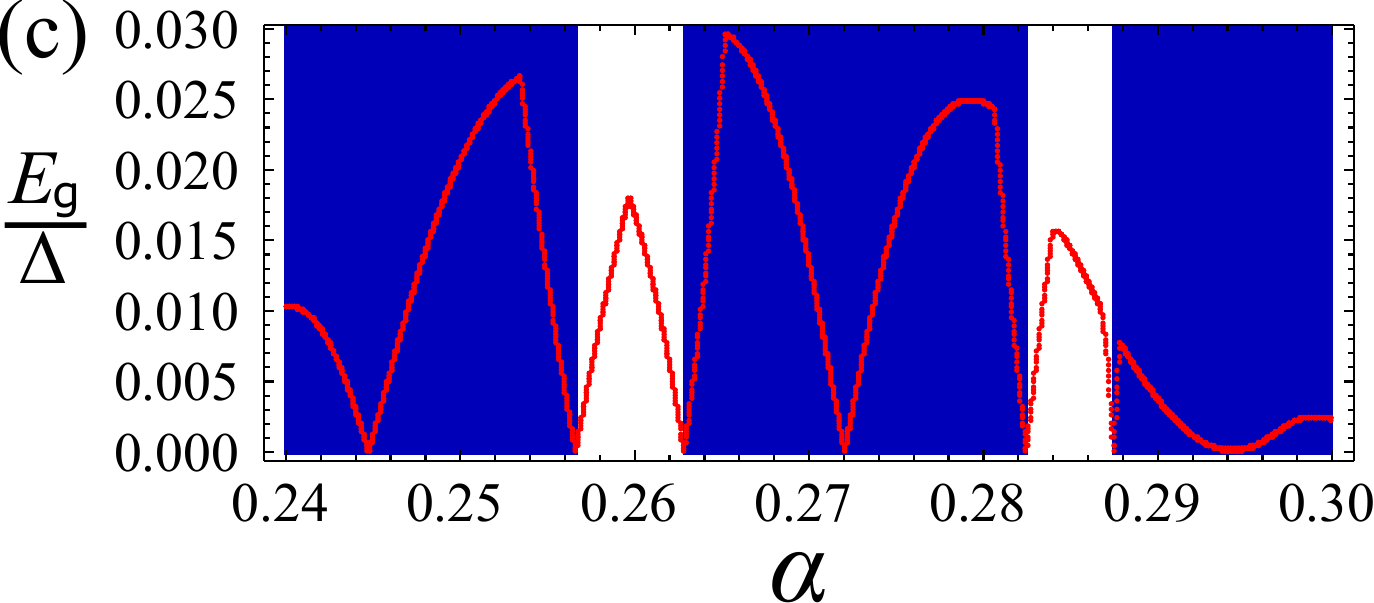}\\
\includegraphics[width=2.1in]{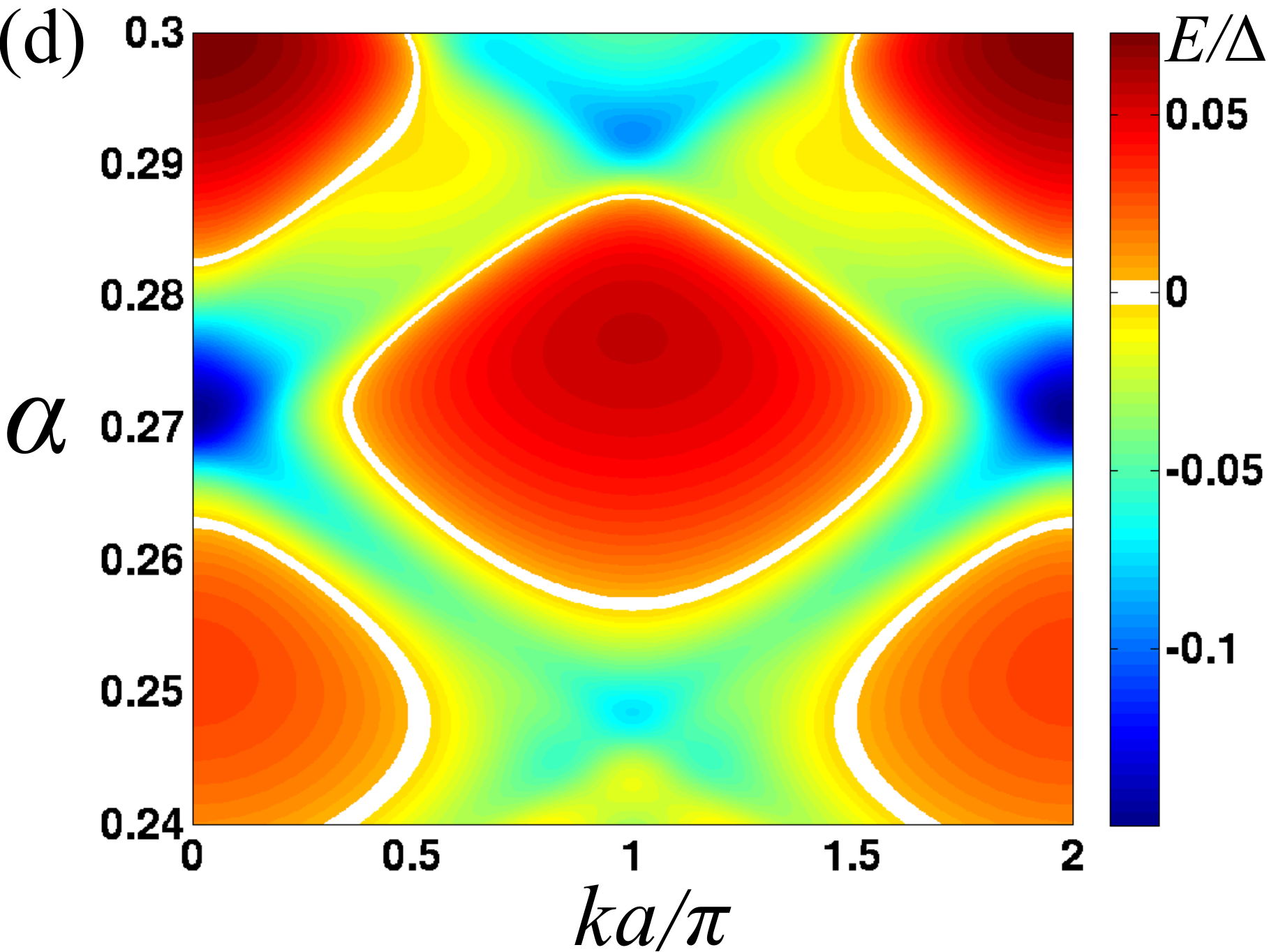}\hfill{}\includegraphics[width=2.1in]{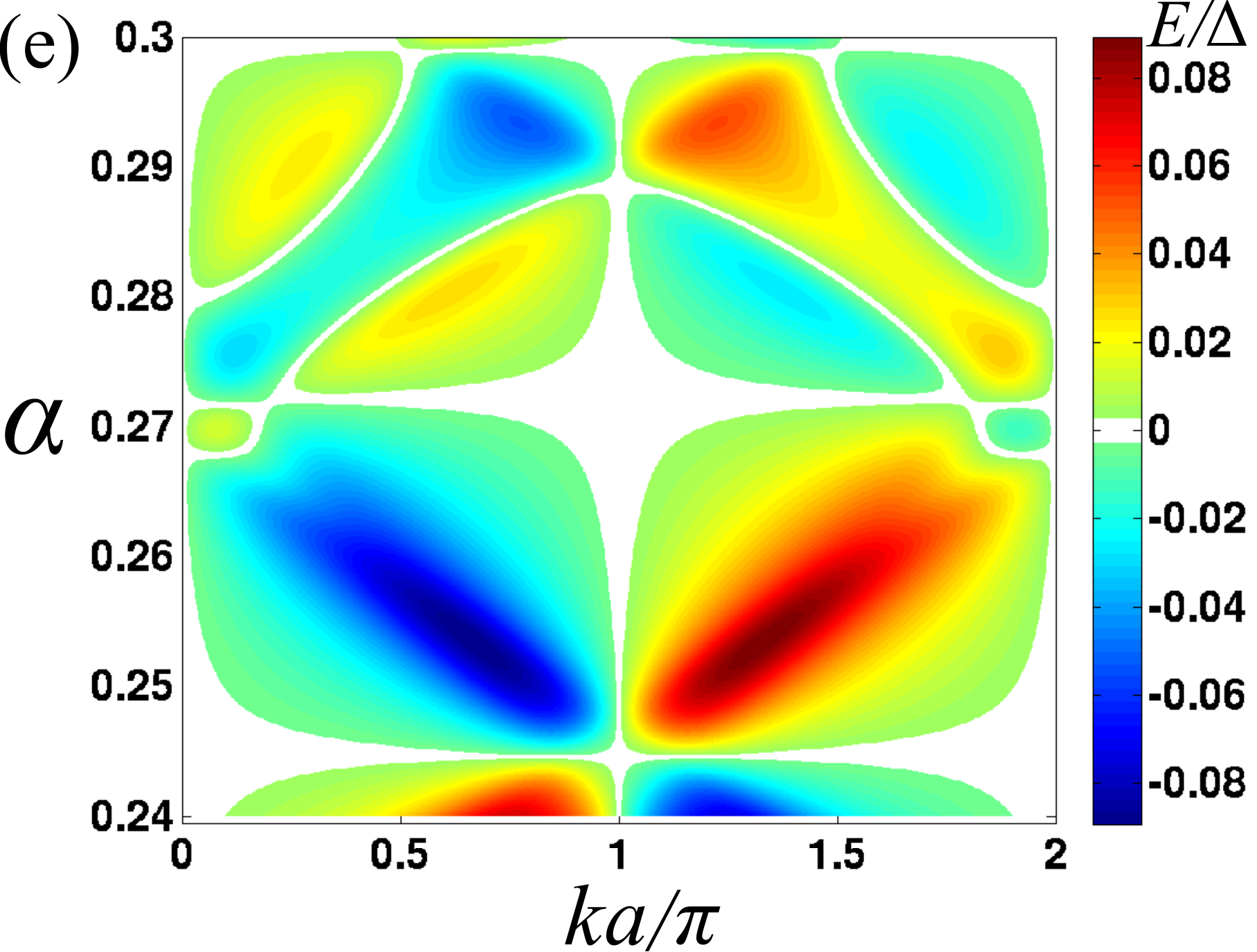}\hfill{} \includegraphics[width=2.1in]{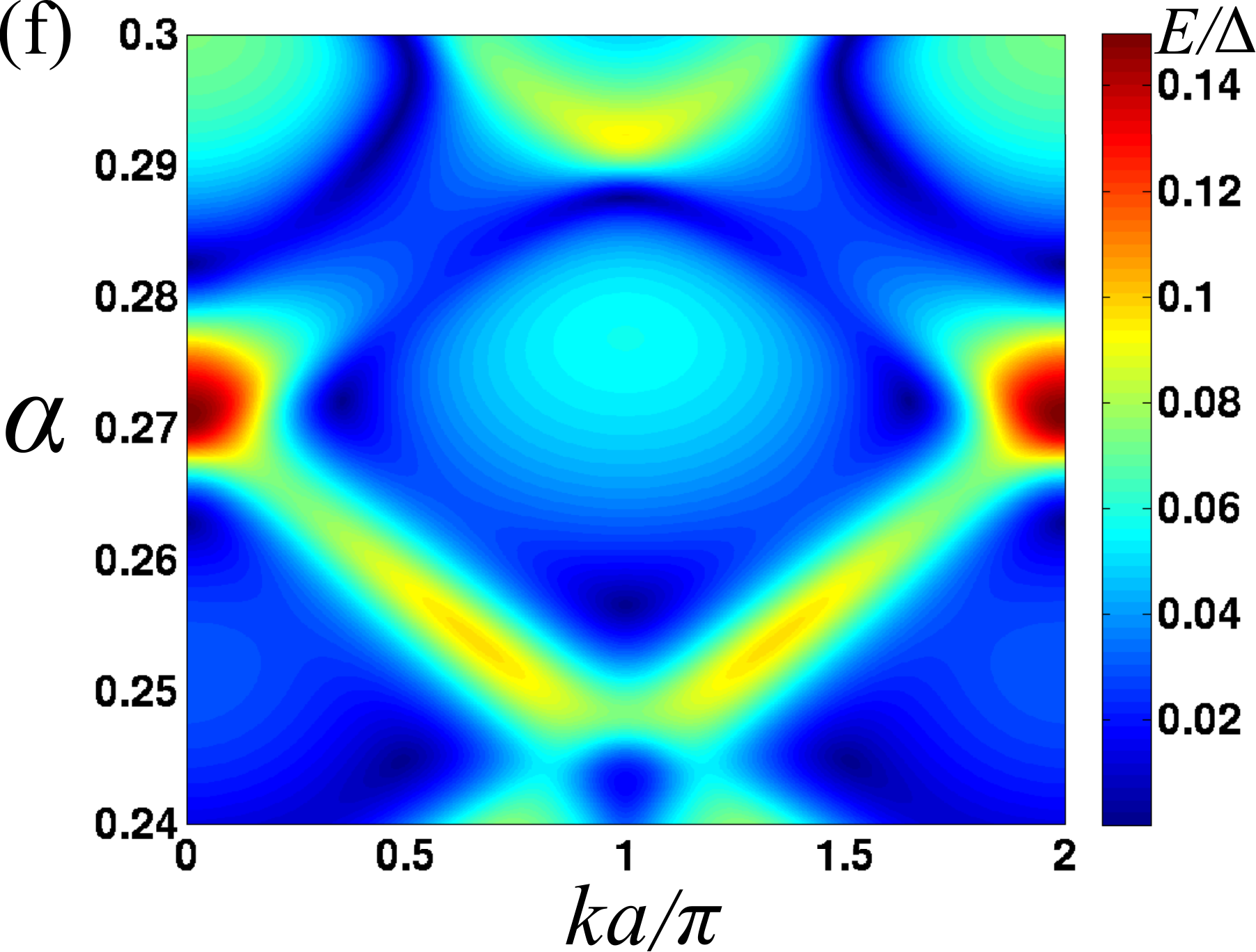}
\caption{(Color online) Quasiparticle excitation gap $E_{g}$ along different line cuts on the phase diagrams:
Panel (a) along the line-cut A in Fig.\,\ref{fig: Fig2}(a) as a function of $J_{0}$ at $k_{F}a=37.2\pi$.
Panel (b) along the line-cut B in Fig.\,\ref{fig: Fig2}(a) as a function of $k_{F}a$ at $J_{0}=1.015$.
Panel (c) along the line-cut C in Fig.\,\ref{fig: Fig2}(b) as a function of $\alpha$ at $k_{F}a=36.8\pi$.
Here the shaded area in (a)-(c) indicates the topologically nontrivial phase as shown in Fig.\,\ref{fig: Fig2}(a)-(b).
To see the location of the gap closing points in Panel (c), we plot the hopping energy spectrum $h(k)$ in panel (d),
the effective pairing energy spectrum $\mathrm{Im}\tilde{\Delta}(k)$ in panel (e), and the whole
energy spectrum of the particle band in panel (f), as a function of $k$ and $\alpha$. \label{fig: Fig3}}.
\end{figure*}

In addition to the topological index (Majorana number), we have also calculated quasiparticle excitation gap as a function of $J_{0}$,
$\alpha$, and $k_{F}a$, see Fig.\,\ref{fig: Fig2}(d) and Fig.\,\ref{fig: Fig3}(a)-(c). One can notice that the closing of the gap is consistent with
the phase diagram shown in Fig.\,\ref{fig: Fig2}(a)-(b). Additionally, one has information regarding the magnitude of the gap deep in the topological phase which is crucial for understanding the stability of the topological phase. Fig.\,\ref{fig: Fig3}(c) shows there also exist gap closing points inside the topological phase, not determined by calculating the topological index. To identify the location of these gap closing points, we plot the hopping energy spectrum $h(k)$, the effective pairing energy spectrum $\mathrm{Im}\tilde{\Delta}(k)$, and the energy spectrum of the particle band as a function of $k$ and $\alpha$ in Fig.\,\ref{fig: Fig3}(d)-(f), respectively. Clearly, for certain values of $\alpha$ both hopping and pairing energies vanish at certain momentum points between $0$ and $\pi$ resulting in the gap closing.
In the limit $J_1\to 0$ it can be shown analytically that these accidental zeros occur when the $k_{F,\pm}$ are commensurate with each other,
i.e. when $\alpha = n\pi/(k_Fa)$, with $n\in\mathbb{N}$. We will discuss the origin of these accidental gap closing points in more details in the next section.

\section{deep $P$ band limit}\label{sec:p-band}

\subsection{Derivation of the Effective Hamiltonian}

We now discuss the deep $p$-band limit assuming that the energy of the $l=\pm1$ states is lower than that of $l=0$, i.e., $J_{0}\ll J_{1}$.
In the limit $J_{1}\sim 1$ and $\alpha^2J_0\ll 1$, the on-site energy $\epsilon\approx\Delta(1-J_{1})+\mathcal{O}(\alpha^{2})$ is close to the midgap~\cite{kim2014b}. Once again, we assume that the $l=\pm1$ states are well separated from the $l=0$ state, and, therefore, $l=0$ states can be integrated out.
As a result, we obtain a tight-binding description for the $p$-bands with the $s$ channel taken into account perturbatively
by allowing for the transitions through intermediate virtual $l=0$ states. Note that there is a significant difference with respect to the calculation in Sec.~\ref{sec:s-band} since there are now two low-energy $p$-bands. Following the same procedure as in the previous section, we obtain the 8 dimensional matrix equation for the deep $p$-band limit:
\begin{equation}
\sum_{j}\mathbf{M}_{p}^{ij}(E)\overline{\Phi}_{j}=0,\label{eq: p_channel_chain_Eq}
\end{equation}
where $\overline{\Phi}_{i}=\left(\overline{\psi}_{i,-1},\,\overline{\psi}_{i,1}\right)^{T}$
is the 8 dimensional spinor for the $p$-channel states. The derivation
of the matrix $\mathbf{M}_{p}^{ij}(E)$ is presented in the Appendix \ref{app:E}. Assuming that $k_Fa \gg 1$ and $\epsilon\rightarrow 0$, the $p$-bands have narrow bandwidth with the center of the bands being close to the midgap. One can then linearize Eqs.\eqref{eq: p_channel_chain_Eq} with respect to $E$: $\mathbf{M}_{p}^{ii}(E)\approx\mathbf{M}_{p}^{ii(0)}-\mathbf{M}_{p}^{ii(1)}\cdot E$, and neglect the energy dependence of the inter-site matrix $\lim_{E\rightarrow0}\mathbf{M}_{p}^{i\neq j}(E)\equiv\mathbf{M}_{p}^{i\neq j}(0)$ by dropping the terms $\mathcal{O}\left(\frac{E}{\Delta\sqrt{k_{F}a}}\right)$ and $\mathcal{O}\left(\frac{\alpha^{2}}{\sqrt{k_{F}a}}\right)$ with $\alpha \ll 1$ (i.e. we will keep henceforth the terms only up to $\mathcal{O}(\alpha^{2})$ and $\mathcal{O}(\alpha/\sqrt{k_{F}a})$.). After some algebra (see Appendix~\ref{app:E} for details), Eq.\,(\ref{eq: p_channel_chain_Eq}) can be written as
\begin{equation}
\sum_{j}H_{p}^{ij}\overline{\Phi}_{j}=E\overline{\Phi}_{i},\label{eq: p_band_linear_chain_Eq}
\end{equation}
where
$H_{p}^{ij}=\left(\mathbf{M}_{p}^{ii(1)}\right)^{-1}\mathbf{M}_{p}^{ij}(0)$.
We then project $H_{p}^{ij}$
onto the local basis of YSR states: $\left(\begin{array}{cccc}
\phi_{1,+}, & \phi_{2,+}, & \phi_{1,-}, & \phi_{2,-}\end{array}\right)^{T}$ where $\phi_{1(2),\pm}$ are the eigen spinors of the single-impurity
bound states with energy $\pm\epsilon_{1(2)}$ in channel 1 ($l=-1$)
and channel 2 ($l=1$). The local basis can be found by solving
the single-site equation $\mathbf{M}_{p}^{ii}(E)\phi=0$ as a special
case of Eq. (\ref{eq: p_channel_chain_Eq}), where the bound state
energies are determined from $\mathrm{Det}\left[\mathbf{M}_{p}^{ii}(E)\right]=0$, see Appendix~\ref{app:F}.
After the projection onto the local basis we obtain the effective Hamiltonian
$\mathcal{H}_{p}(k)$ describing the two coupled bands of the YSR chain in the deep-$p$~band limit.
%

\begin{figure}
\hfill{}\includegraphics[width=3.2in]{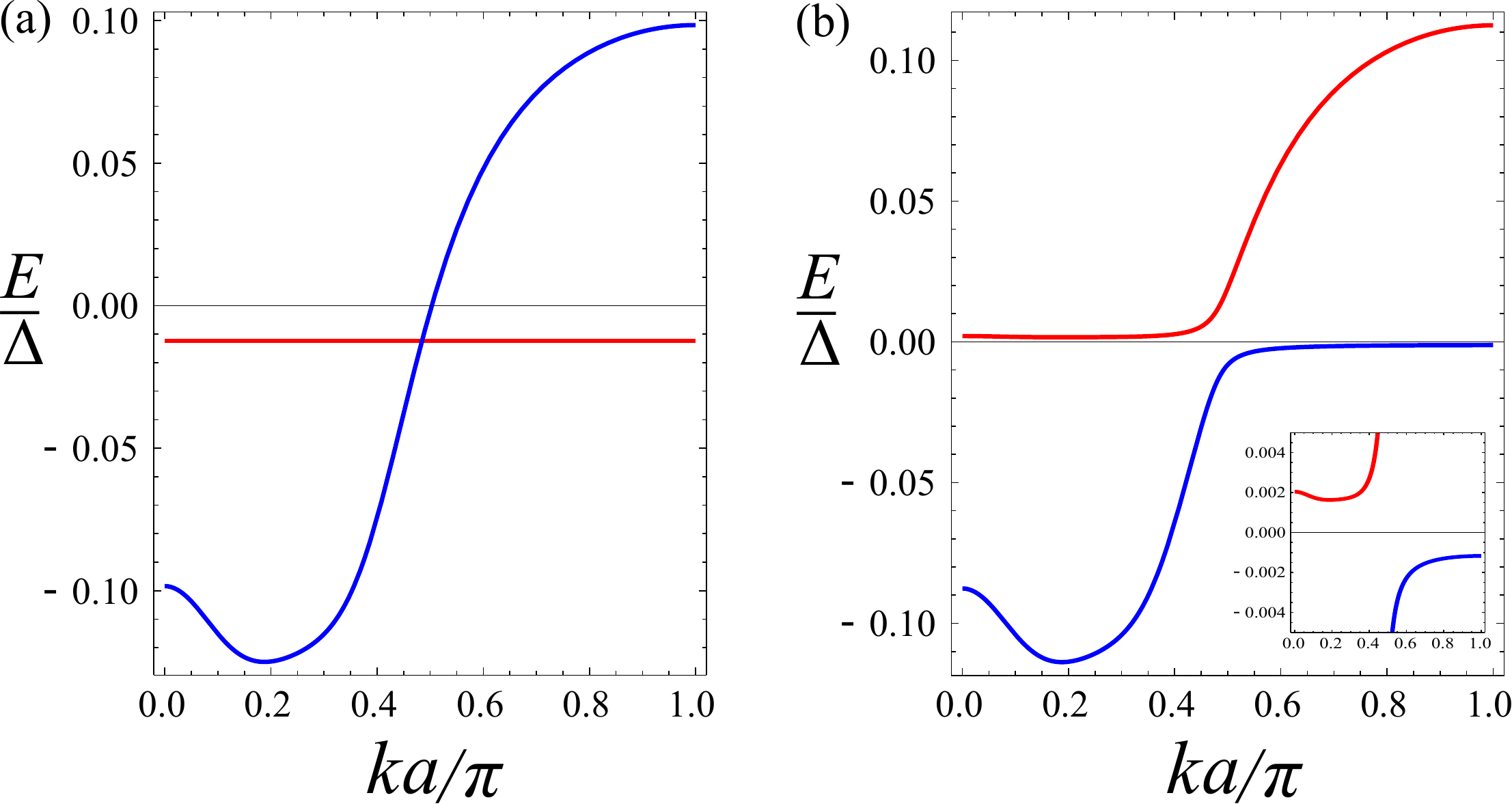}\hfill{}
\caption{(Color online) The dependence of the normal-state energy spectrum (i.e. $\Delta_{ij}=0$) on momentum $k$ for (a) $\alpha J_{0}=0$, $J_{1}=1.0125$, $k_{F}a=37.5\pi$, $\xi_0=2a$; and (b) $\alpha=0.3$, $J_0=0.4$, $J_{1}=1.0125$, $k_{F}a=37.5\pi$, $\xi_0=2a$. The normal-state spectrum consists of heavy-fermion and light-fermion bands which are hybridized by the SOC. The zoom-in figure of panel (b) near the Fermi level is shown in the inset. \label{fig: Fig4}}
\end{figure}

Simple analytical expressions for $\mathcal{H}_{p}(k)$ can be obtained
when the impurity spins are polarized normal to the plane, i.e. along
$\hat{\mathbf{z}}$ direction.
In the case that the impurities
are polarized in $\hat{\mathbf{z}}$ direction, transforming the effective
tight-binding Hamiltonian $\mathcal{H}_{p}(i,j)$ to momentum space,
we obtain a two-band BdG Hamiltonian,
\begin{align}\label{eq: p_band_H}
\frac{\mathcal{H}_{p}^{\hat{z}}(k)}{\Delta}=\left(\begin{array}{cccc}
h_{11}(k) & h_{12}(k) & \tilde{\Delta}_{11}(k) & \tilde{\Delta}_{12}(k)\\
h_{21}(k) & h_{22}(k) & \tilde{\Delta}_{21}(k) & \tilde{\Delta}_{22}(k)\\
\tilde{\Delta}_{11}^{*}(k) & \tilde{\Delta}_{21}^{*}(k) & -h_{11}(k) & -h_{21}(k)\\
\tilde{\Delta}_{12}^{*}(k) & \tilde{\Delta}_{22}^{*}(k) & -h_{12}(k) & -h_{22}(k)
\end{array}\right).
\end{align}
The coefficients here satisfy the following properties: $h_{ij}(k)=h_{ij}(-k)$, $\tilde{\Delta}_{ij}(-k)=-\tilde{\Delta}_{ij}(k)$
and, therefore, $\tilde{\Delta}_{ij}(p)=0$ at $p=0,\ \pi/a$. The effective hopping energies include both intra-channel hopping
\begin{align}
h_{11}(k)&\simeq\epsilon_{1}+\frac{1}{2}\left[I_{0,+}(k)+I_{0,-}(k)\right],\\
& \stackrel{k \rightarrow 0}{\approx} h_{11}^{(0)} + h_{11}^{(2)}k^2\nonumber\\
h_{22}(k)&\simeq\epsilon_{2}+\frac{1}{2}\left[I_{0,+}(k)+I_{0,-}(k)\right],\\
& \stackrel{k \rightarrow 0}{\approx} h_{22}^{(0)} + h_{22}^{(2)}k^2.\nonumber
\end{align}
and inter-channel hopping
\begin{align}
h_{12}(k)=h_{21}(k)&\simeq\frac{1}{2}\left[I_{2,+}(k)+I_{2,-}(k)\right],\\
 \stackrel{k \rightarrow 0}{\approx} & h_{12}^{(0)} + h_{12}^{(2)}k^2\nonumber.
\end{align}
with the on-site energies
\begin{align}
\epsilon_{1} & \simeq\frac{1-J_{1}}{J_{1}}+\frac{\alpha^{2}J_{0}(2-J_{1}+J_{0})}{J_{1}(1+J_{0})^{2}},\\
\epsilon_{2} & =\frac{1-J_{1}}{J_{1}}.
\end{align}
The effective p-wave pairing also contains both intra-channel pairing
\begin{align}
\tilde{\Delta}_{11}(k) & \simeq\frac{i}{2}\left[K_{1,+}(k)-K_{1,-}(k)\right]\\
 & -\frac{i\alpha J_{0}}{1+J_{0}}\left[K_{1,+}(k)+K_{1,-}(k)\right],\nonumber \\
& \stackrel{k \rightarrow 0}{\approx} \Delta_{11}^{(1)}k.\nonumber\\
\tilde{\Delta}_{22}(k)&\simeq\frac{i}{2}\left[K_{3,+}(k)-K_{3,-}(k)\right],\\
 &\stackrel{k \rightarrow 0}{\approx}  \Delta_{22}^{(1)}k\nonumber.
\end{align}
and inter-channel pairing
\begin{align}\label{eq: inter_pairing_p_band}
\tilde{\Delta}_{12}(k)=\tilde{\Delta}_{21}(k) & \simeq\frac{i}{2}\left[K_{1,+}(k)-K_{1,-}(k)\right] \\
 & -\frac{i\alpha J_{0}}{2\left(1+J_{0}\right)}\left[K_{1,+}(k)+K_{1,-}(k)\right]\nonumber.\\
& \stackrel{k \rightarrow 0}{\approx} \Delta_{12}^{(1)}k\nonumber.
\end{align}
The coefficients of the small $k$ expansions are explained in the Appendix~\ref{app:G}.

In order to understand the physics described by the Hamiltonian~\eqref{eq: p_band_H}, we first discuss the effect of SOC on the normal-state band structure (i.e. $\Delta_{ij}=0$). The spectrum for the two bands reads
\begin{equation}
\frac{E_{\pm}^{N}(k)}{\Delta}=\frac{1}{2}\left[h_{11}(k)+h_{22}(k)\pm\sqrt{4h_{12}^{2}(k)+\left(\delta\epsilon_{12}\right)^{2}}\right]
\end{equation}
where $\delta\epsilon_{12}=\epsilon_{1}-\epsilon_{2}$.
As shown in the Appendix~\ref{app:A}, to leading order in $1/\sqrt{k_{F}a}$,
$I_{0,\lambda}(k)\approx I_{2,\lambda}(k)$. Hence $h_{12}$ is approximately the same as $h_{11}$ and $h_{22}$. In the absence of SOC, $\delta\epsilon_{12}$ vanishes, and the band structure
is characterized by a heavy-fermion band $E_{1}^{N}\approx(1-J_{1})/J_{1}$ crossing with
a dispersive light-fermion band $E_{2}^{N}\approx(1-J_{1})/J_{1}+\left[I_{0,+}(k)+I_{0,-}(k)\right]$
with the bandwidth doubled compared to $s$-band, as shown in Fig.~\ref{fig: Fig4}(a). The on-site orbital structure of these two bands are symmetric(light) and anti-symmetric(heavy) combinations of $l=\pm 1$ states. The physical origin of these orbital structures reflects the degeneracy due to the isotropic magnetic potential and the asymptotically equal hopping amplitudes.

In the presence of SOC and a finite $s$ channel coupling (i.e. $J_{0}\neq0$
and $\alpha\neq0$) $\delta\epsilon_{12}$ becomes finite, which induces
a hybridization gap between the two bands leading to an avoided level crossing,
as shown in Fig.\,\ref{fig: Fig4}(b). The induced hybridization
gap gives rise to an interesting feature in the topological phase diagram
as discussed below.

\subsection{Topological Properties}
The topological phase diagram in the deep~$p$-band limit (\ref{eq: p_band_H})
involves two bands which are hybridized by the SOC. As a
result, it exhibits a more intricate dependence on the parameters
compared with the deep $s$-band limit. In order to calculate the
$Z_{2}$ topological invariant $\mathcal{M}$, we use the method
developed for the multiband semiconductor nanowire
system \cite{1DwiresLutchyn2}:
$\mathcal{M}=\mathrm{sgn}\left[\mathrm{Pf}B(0)\mathrm{Pf}B(\pi/a)\right]=\pm1$,
where the antisymmetric matrix
$B(p)=\mathcal{H}_{p}^{\hat{z}}(p)\tau_{x}$ is calculated at the
particle-hole invariant points: $p=0,\ \pi/a$.  For the two-band
system, the corresponding expression for the Pfaffian is
$\mathrm{Pf}B(p)=h_{12}(p)h_{21}(p)-h_{11}(p)h_{22}(p)$.

\begin{figure}
\center
\includegraphics[height=1.75in]{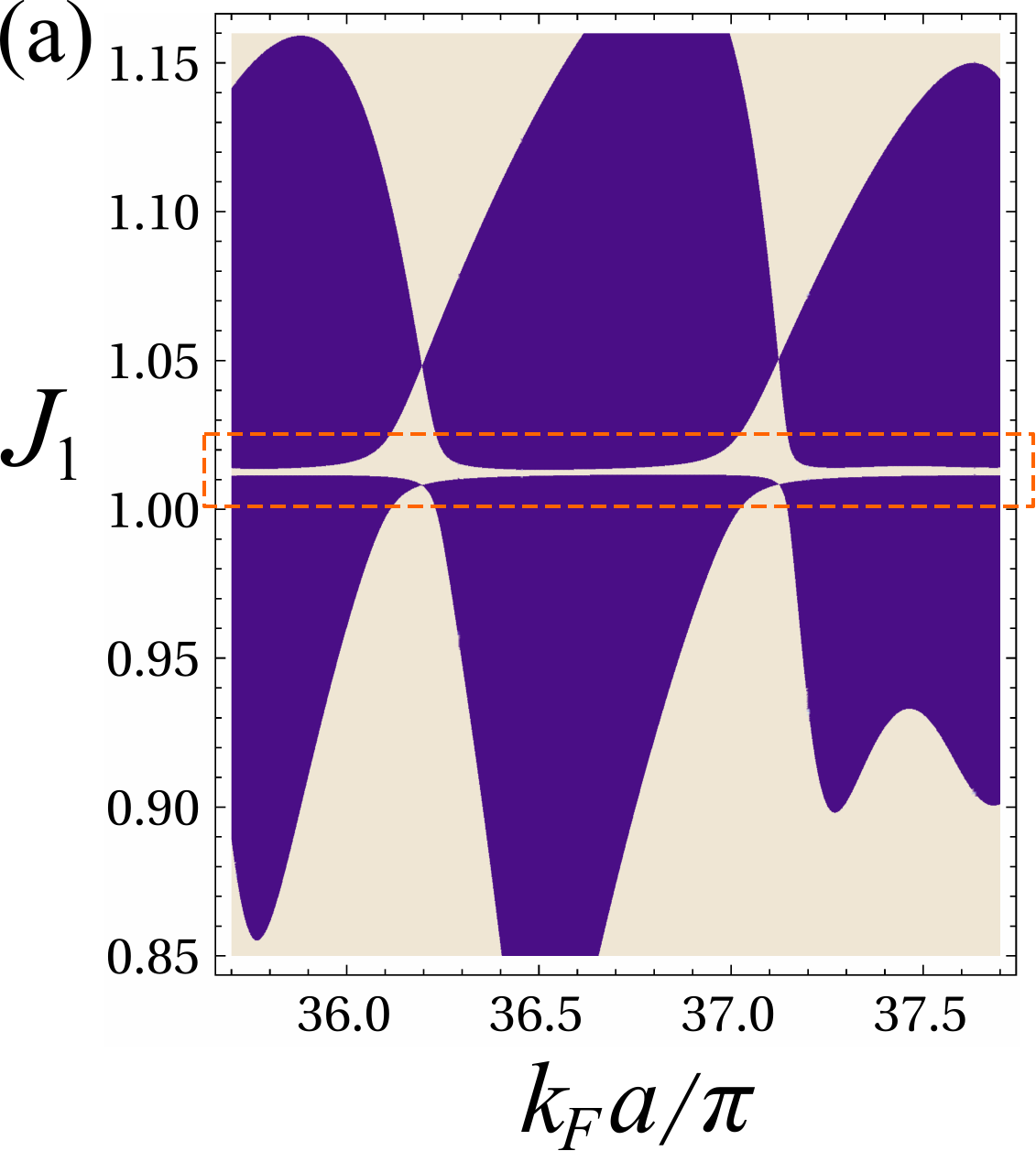}\hfill{}\,\,\,\,\,\includegraphics[height=1.9in]{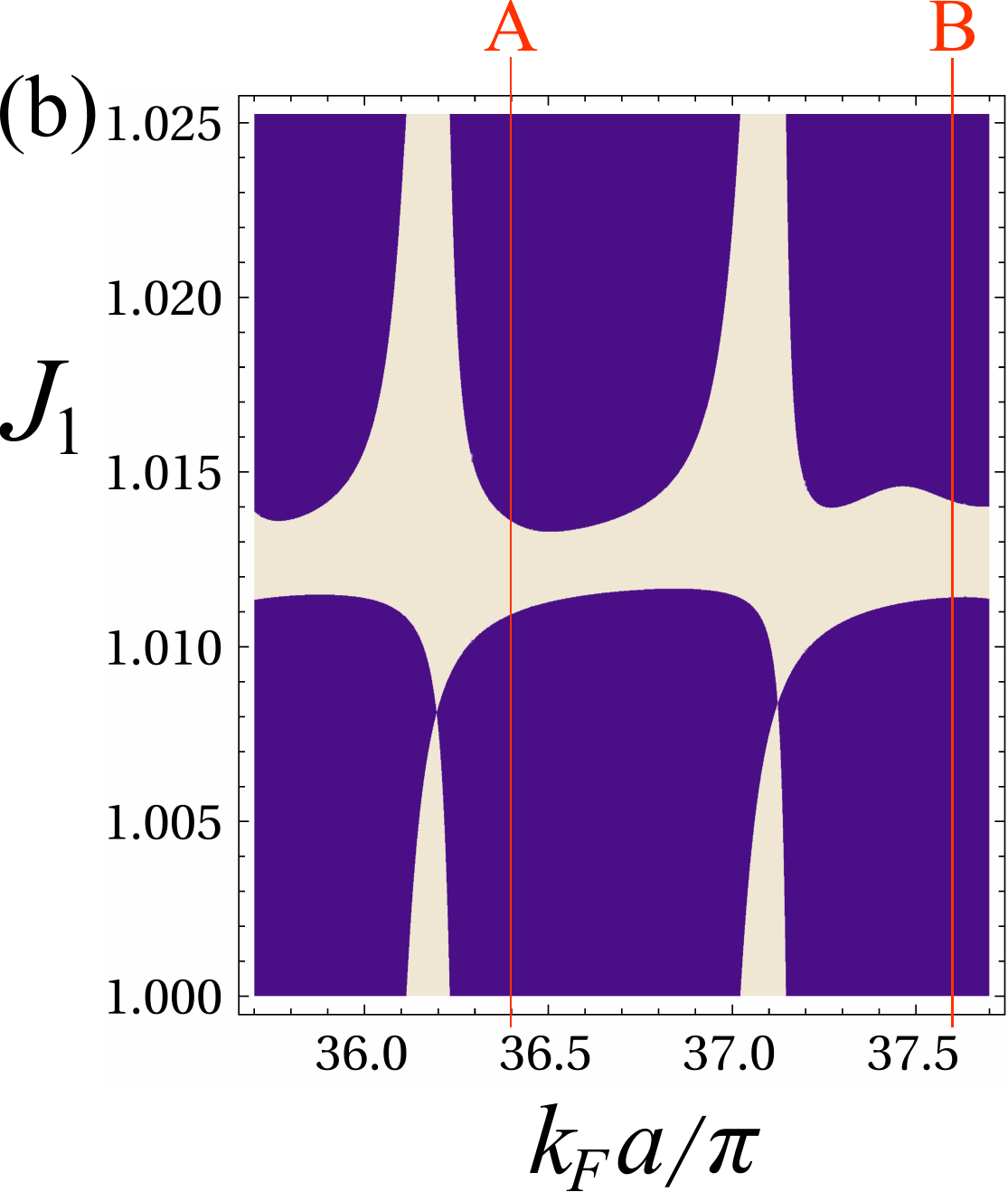} \\
\includegraphics[height=1.75in]{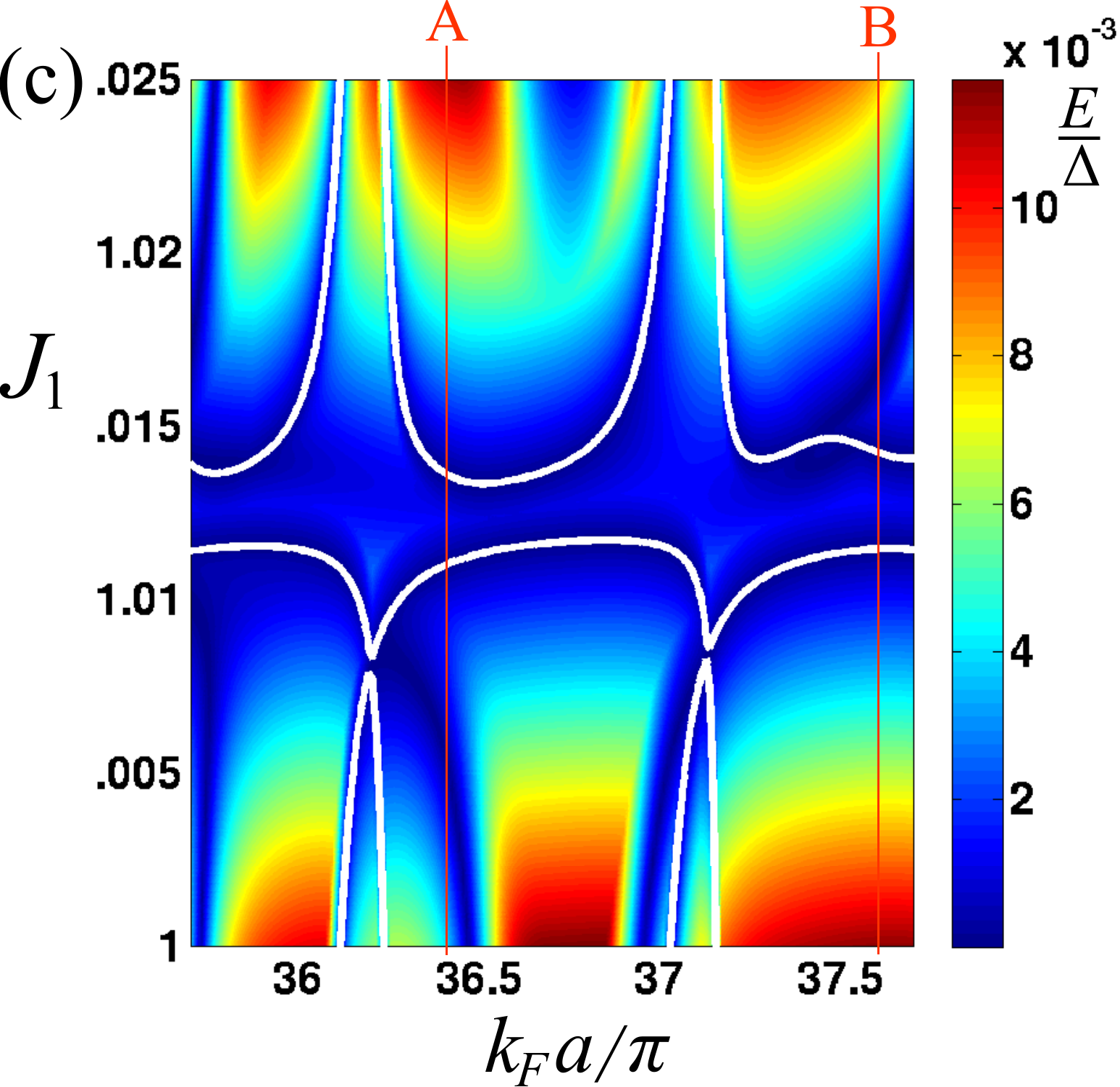}\hfill{}\,\,\,\includegraphics[height=1.65in]{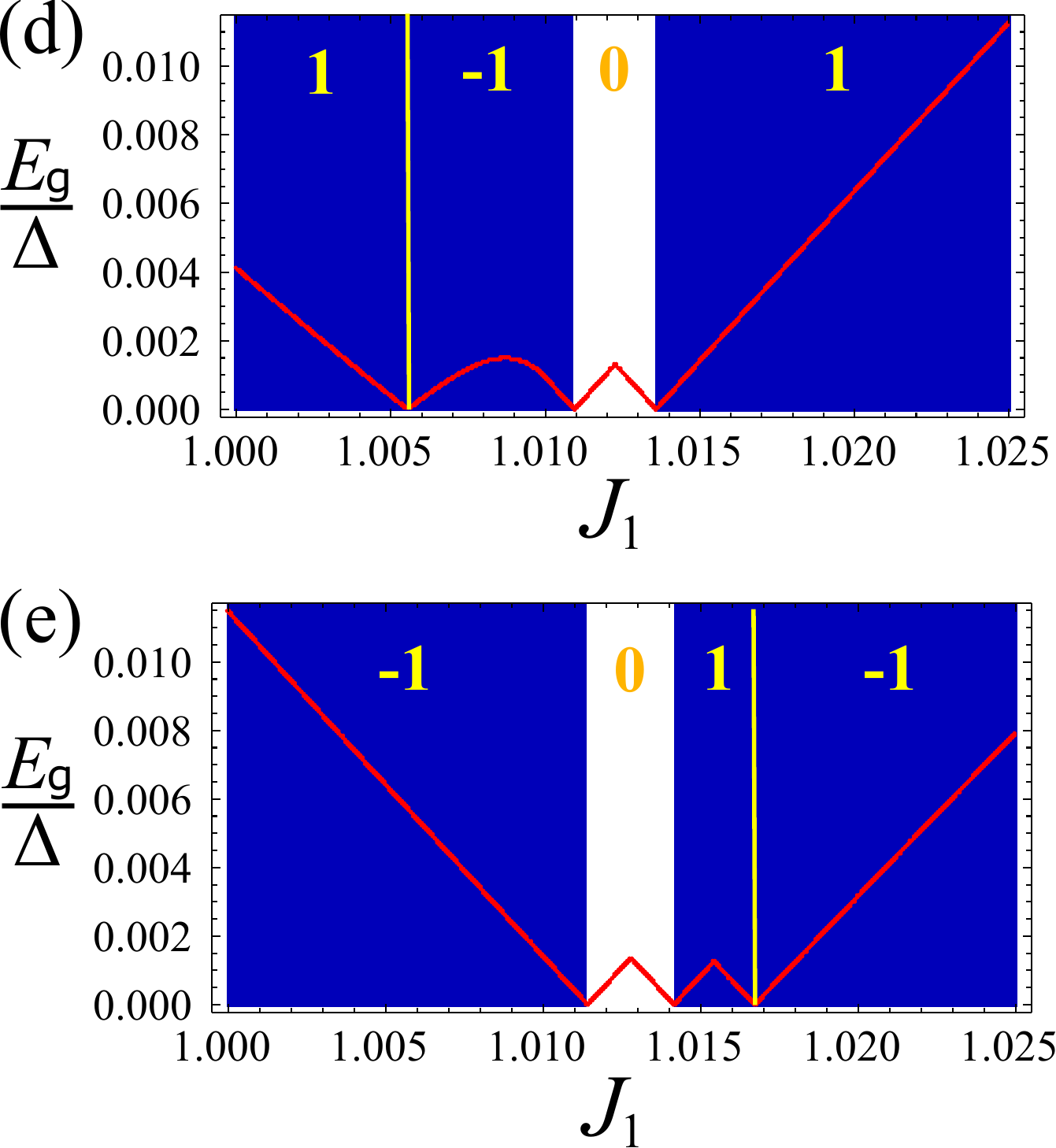}
\caption{(Color online)
(a) Topological phase diagram in the $(k_Fa, J_1)$ plane in the deep~p-band limit, for the case in
    which $\mathbf{S}\parallel\hat{\mathbf{z}}$, $J_{0}=0.4$, $\alpha=0.3$, and $\xi_0=2a$.
(b) Enlargement of the topological phase diagram shown in (a) around the region surrounded by the dashed line rectangle.
(c) Calculated quasiparticle excitation gap for the parameter regime in the phase diagram (b) with the phase boundary indicated by white line.
(d) The quasiparticle excitation gap on the line-cut A in panel (b) and (c) near the re-entrance region at $k_{F}a=36.4\pi$.
(e) The quasiparticle excitation gap on the line-cut B in panel (b) and (c) near the re-entrance region at $k_{F}a=37.6\pi$. Here the shaded area in (d) and (e) indicates the topologically nontrivial phase as shown in panel (b). The integer numbers -1, 0, 1 shown in (d) and (e) are the winding numbers calculated for each gapped phase. \label{fig: Fig5}}
\end{figure}

The topological phase diagram for the effective Hamiltonian (\ref{eq:
p_band_H}) as a function of $J_{1}$ and $k_{F}a$ is shown in
Fig.~\ref{fig: Fig5}~(a).
We immediately notice that in the deep~p-band limit there is
a large fraction of the $(J_1, k_Fa)$ plane in which the chain is in a topological phase.
This is somewhat surprising given that in the deep~p-band we have two bands close in
energy and one could expect that the hybridization of the two bands could lead
for most values of $J_1$ and $k_Fa$ to a situation in which the Fermi energy
intersects the bands an even number of times.
The reason why this is not the case is evident from the plots of the two bands
for a typical situation shown in Fig.~\ref{fig: Fig4}: one can see that the result
of the hybridization of the bands leads to the formation of the light- and heavy-fermion bands.
As a consequence, for most values of $J_1$ and $k_Fa$ the Fermi energy crosses only once
the light-fermion band. On the other hand, we expect that when $J_1$ and $k_Fa$
are such that the Fermi energy is very close to the heavy band the chain should be
in a topologically trivial phase. Given the flatness of the heavy band we expect that this will
happen only for a very small range of values of $J_1$.
Indeed, a close inspection of the topological phase diagram of Fig.~\ref{fig: Fig5}~(a)
shows that for $J_1$ very close to 1 there is a narrow region,
highlighted in the figure by a rectangular box, in which the chain is in a topologically
trivial phase.

To understand the re-entrance region from the topological phase to the trivial
phase when $J_{1}\sim 1$, in Fig.~\ref{fig: Fig5}~(b) we show a zoom-in of the
topological phase diagram in the region surrounded by the rectangular box in
Fig.~\ref{fig: Fig5}~(a). In the zoom-in plot the presence of a
topologically trivial phase is clearly visible.
In this region, due the fact that  $J_1 \rightarrow 1$
the effective chemical potential is close to zero so that
the heavy-fermion band becomes important: in this situation the chemical potential
lies inside the hybridization gap, as shown in Fig.~\ref{fig: Fig4}~(b), which
results in the re-entrance to the trivial phase.
Outside this narrow range of values of $J_1$ the heavy-fermion band is
either completely empty or filled and the chemical potential crosses the light-fermion band an odd number of times,
thus, there is an odd number of Majorana zero modes per end
originating from the light-fermion band. Due to this
interplay, the two YSR bands do not annihilate each other (except for
the small region close to $J_1 \rightarrow 1$) which is a peculiar
feature of this model.

In order to understand the stability of the topological phase and to
corroborate the results discussed above, we also compute the quasiparticle
excitation gap $E_g$, see Fig.~\ref{fig: Fig5}(c). Figure~\ref{fig:
Fig5}(d)-(e) plot the value of $E_g$ along the two line-cuts on the
phase diagram near the re-entrance region. One can see that the
quasiparticle gap closing is consistent with the topological phase
diagram in Fig.~\ref{fig: Fig5}(b).  We once again find that SOC
controls the magnitude of the quasiparticle gap and, as such, is
crucial for the stability of the topological phase.

In addition to the quasiparticle gap closing at the topological phase
boundary, one can notice that there are also points where the gap vanishes
in the topological phase, see Fig.\ref{fig: Fig5}(d) at $J_1\sim 1.005$ and Fig.\ref{fig: Fig5}(e) at $J_1\sim 1.017$. We now
investigate in detail the two regions across these points using an
additional symmetry of our effective Hamiltonian~\eqref{eq:
p_band_H}. In addition to the particle-hole symmetry
$\mathcal{P}=\tau_x\mathcal{K}$ where $\mathcal{K}$ refers to
complex conjugation, our effective spinless Hamiltonian also has a
pseudo-time reversal symmetry $\mathcal{T}=\mathcal{K}$. Using these
two symmetries, one can construct another symmetry - chiral symmetry
$\mathcal{S=TP}=\tau_x$ which anticommutes with the
Hamiltonian~\eqref{eq: p_band_H}. Thus, the Hamiltonian~\eqref{eq:
p_band_H} belongs to the BDI symmetry class~\cite{altland1997,
schnyder2008, kitaev2008} which is characterized by the integer
invariant $\mathcal{W}$ and supports multiple spatially-overlapping
Majorana zero modes~\cite{Cheng2010}. In order to calculate the
topological index $\mathcal{W}$, it is convenient to transform the
Hamiltonian~\eqref{eq: p_band_H} into a chirality basis using a unitary
transformation $\mathcal{U}=e^{-i\frac{\pi}{4}\tau_y}$ which converts
the Hamiltonian to the off-diagonal form: \be
\mathcal{U}\mathcal{H}^{\hat{z}}_p(k)\mathcal{U}^\dagger=\left(\begin{array}{cc}
0 & A(k)\\ A^\dag(k) &0
\end{array}\right).  \ee Then, the winding number (i.e. the number of
Majorana modes per each end) can be calculated by introducing a
complex variable $z(k)=\det[A(k)]/|\det[A(k)]|$, and calculating the
integral \be \mathcal{W}=-\frac{i}{\pi}\int_{k=0}^{k=\pi}
\frac{dz(k)}{z(k)}, \ee Using this analysis we find that, for example,
the phases at $J_1=1.001$ and $J_1=1.007$ in Fig.~\ref{fig: Fig5}(d)
have different winding numbers $\mathcal{W}(J_1=1.001)= 1$ and
$\mathcal{W}(J_1=1.007)= -1$. Thus, gap closing between these two
regions corresponds to the transition between $\mathcal{W}=\pm 1$. The same argument holds for $J_1=1.015$. Thus, accidental gap closing points inside of the topological or non-topological phases are not really accidental but represent the change of the winding number by an even integer.

The analysis above relies on the chiral symmetry. However, in realistic systems the chiral
symmetry can be easily broken by allowing, for example, for a generic
direction of magnetic chain polarization (i.e. along $y$-axis). The
precise magnitude for the Majorana splitting energy, which is
important for tunneling transport measurements, depends on the details
of the chiral-symmetry-breaking perturbations, and we refer a reader
to Refs.\cite{niu2012,LiJ2014,Dumitrescu2014,Heimes15,Hui2015} for
more details. As a consequence, the topological phases identified
by the parity of the topological index are expected to be much more
robust and this is the reason that our analysis has been focused mostly
on characterizing the dependence of such index on the parameters of the system.

\section{conclusions}\label{sec:conclusions}

\begin{figure}
\includegraphics[height=1.77in]{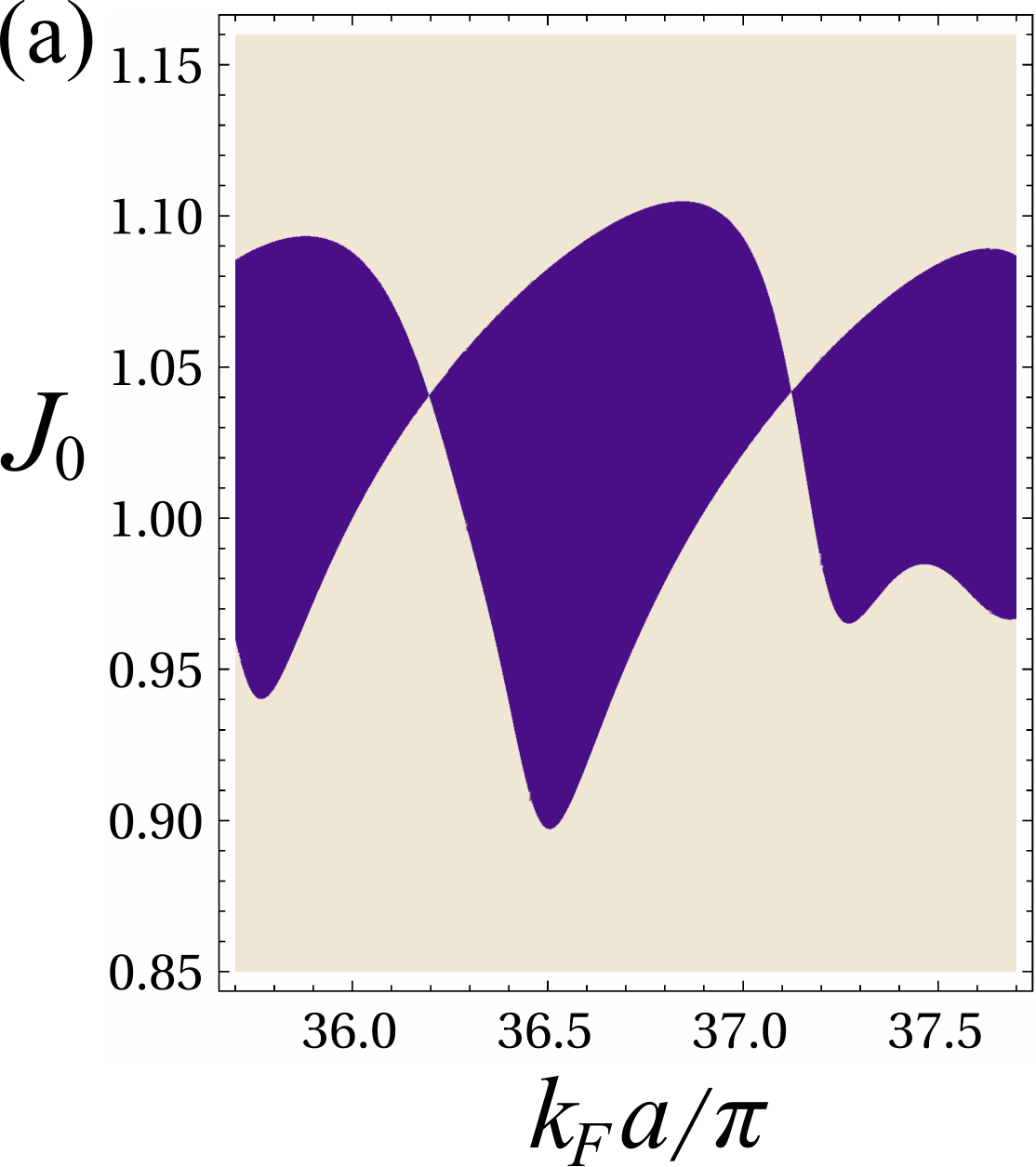}\hfill{} \hfill{}\includegraphics[height=1.77in]{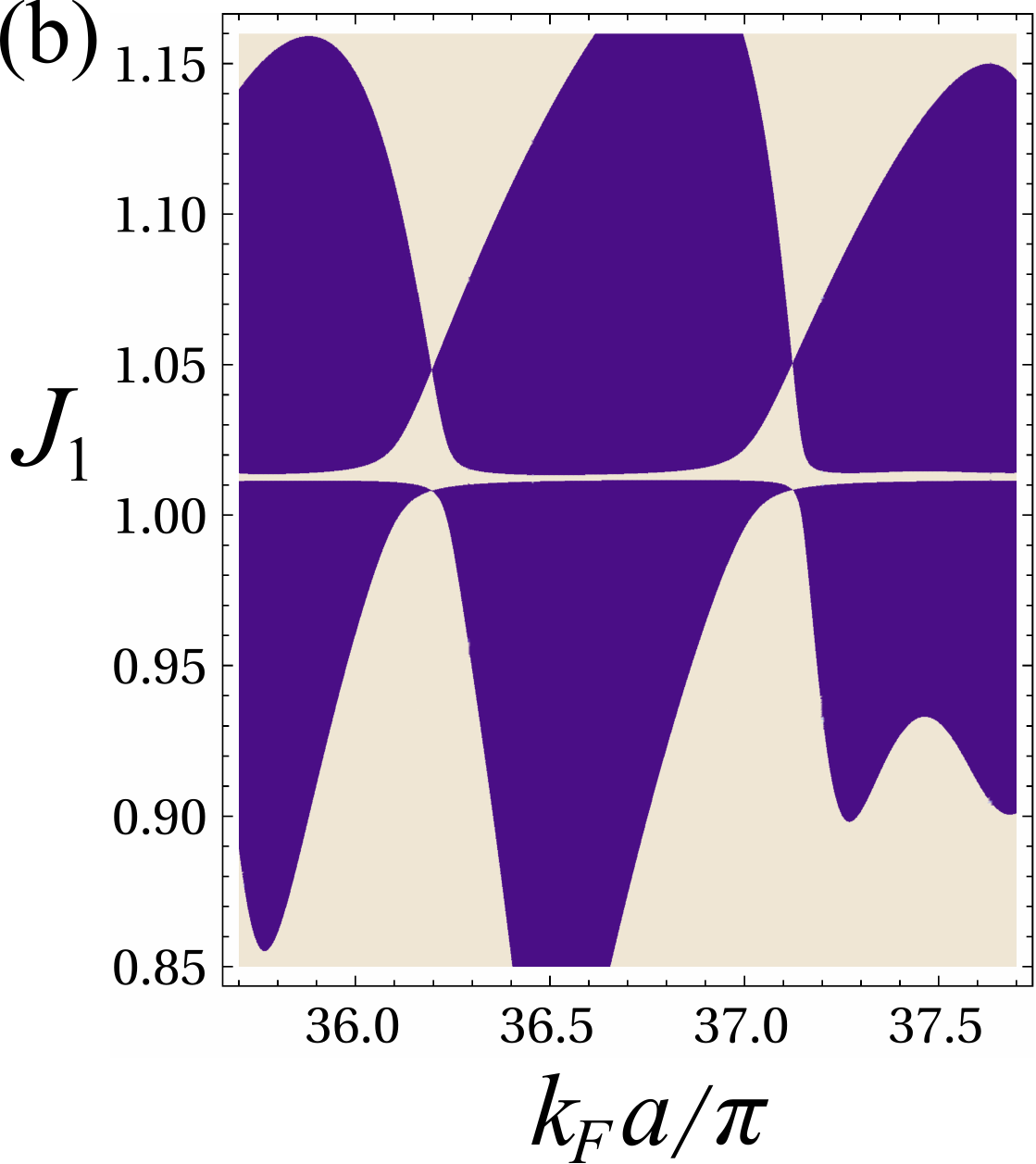}
\caption{(Color online) Comparison of the topological phase diagram in (a) the
$s$ band system for $\alpha=0.3$, $J_{1}=0.4$, and $\xi_0=2a$;
and in (b) the $p$ band system for $\alpha=0.3$, $J_{0}=0.4$, and $\xi_0=2a$. \label{fig: Fig6}}
\end{figure}

We have studied the topological properties of a chain of magnetic impurities placed on
the surface of an s-wave superconductor with Rashba spin-orbit coupling taking into account
the presence of multiple scattering channels, in the limit in which the states induced by isolated impurities
are well described as Yu-Shiba-Rusinov states and the distance $a$  between the impurities forming the chain
is such that $k_Fa\gg 1$.
The inclusion of multiple angular momentum scattering channels, $l$, implies that for the chain of YSR states
we have multiple bands. We have shown that the multiband character of the bands strongly affects the topological
properties of the chain for the case when the lowest energy bands are the ones arising from the hybridization
of YSR states with $|l|>0$. Considering the lowest $l=0$ and $|l|=1$ channels we have obtained the topological
phase diagram in the deep~$s$-band and deep~$p$-band limits. Our results show, somehow unexpectedly, that even in the
deep~$p$-band limit there is a large region of parameters space in which the chain is in a topological phase.
Moreover, we find that even though the deep $p$-band case involves two bands, and it is not obvious a priori that these two bands do not ``annihilate" each other, this limit seems to be more favorable for the observation of Majorana zero modes.
This can be seen from the comparison, Fig.~\ref{fig: Fig6}, of the topological phase diagram obtained
in the  deep~$s$-band and deep~$p$-band limit: we see that, for the same range of values of $k_Fa$
and relevant coupling constants ($J_0$ in the deep~$s$-band, $J_1$ in the  deep~$p$-band limit) the
phase space where the chain is in a topologically non-trivial state is larger in the
deep~$p$-band limit than in the deep~$s$-band limit.
We have also characterized the stability of the topological states by computing the quasiparticle excitation gap.

\newpage

Our results have important implications for the ongoing experimental search for Majorana zero modes in this system. The interatomic spacing in Ref.~\cite{nadjperge2014} is of the order of the Fermi wave length ($k_Fa \sim 1$), in which case direct tunneling between iron atoms needs to be included, whereas our calculation assumes $k_Fa \gg 1$. However, at the qualitative level one can already draw a number of conclusions. As $k_Fa$ decreases the bandwidth of subgap YSR bands becomes larger and, thus, the mixing between them becomes even more important. Therefore, it is conceivable that the subgap features reported recently by Yazdani et al.~\cite{YazdaniAPS2015} might correspond to $l>0$ YSR bands. In order to understand these additional subgap features within the framework of our model, one would need to systematically measure the spectrum of the YSR states of single- and two-atom structures~\cite{kim2014b} to identify the angular momentum channels that are the most relevant for the one-dimensional chains. Another important feature that we predict is that SOC allows one to tune the effective chemical potential for the YSR bands and drive the topological phase transition using an external magnetic field. As such, large SOC might be helpful for the manipulation of the Majorana zero modes in this system. At the theoretical level, it would be interesting to establish the correspondence between our results involving classical magnetic impurities with the more microscopic multi-orbital Anderson model for magnetic impurities.

\begin{acknowledgments}
We acknowledge stimulating discussions with B. A. Bernevig, P. M. R. Brydon, E. Gaidamauskas and L. Glazman. JZ and ER acknowledge support from ONR, Grant No. ONR-N00014-13-1-0321. YK acknowledges support from Samsung Scholarship. RL wishes to acknowledge the hospitality of the Aspen Center for Physics and support under NSF Grant \#1066293.
\end{acknowledgments}
\appendix

\begin{widetext}

\section{Analytic Expressions of Integrals $I_{l,\lambda}$ and $K_{l,\lambda}$}\label{app:A}

In this Appendix, we provide analytic expressions of the integral functions defined in Eq.~(\ref{eq:if1})
and Eq.~(\ref{eq:if2}):
\begin{align}
I_{0,\lambda}(x;E)&=\frac{-\Delta\gamma_{\lambda}}{\sqrt{\Delta^{2}-E^{2}}}\mathrm{Re}\left[J_{0}\left((k_{F,\lambda}+i\zeta_{\lambda}^{-1})|x|\right)+iH_{0}\left((k_{F,\lambda}+i\zeta_{\lambda}^{-1})|x|\right)\right]\\
K_{0,\lambda}(x;E)&=\gamma_{\lambda}\mathrm{Im}\left[J_{0}\left((k_{F,\lambda}+i\zeta_{\lambda}^{-1})|x|\right)+iH_{0}\left((k_{F,\lambda}+i\zeta_{\lambda}^{-1})|x|\right)\right]\\
I_{1,\lambda}(x;E)&=-\mathrm{sgn}[x]\frac{i\Delta\gamma_{\lambda}}{\sqrt{\Delta^{2}-E^{2}}}\mathrm{Re}\left[J_{1}\left((k_{F,\lambda}+i\zeta_{\lambda}^{-1})|x|\right)-iH_{-1}\left((k_{F,\lambda}+i\zeta_{\lambda}^{-1})|x|\right)\right]\\
K_{1,\lambda}(x;E)&=\mathrm{sgn}[x]i\gamma_{\lambda}\mathrm{Im}\left[J_{1}\left((k_{F,\lambda}+i\zeta_{\lambda}^{-1})|x|\right)-iH_{-1}\left((k_{F,\lambda}+i\zeta_{\lambda}^{-1})|x|\right)\right]\\
I_{2,\lambda}(x;E)&=\frac{\Delta\gamma_{\lambda}}{\sqrt{\Delta^{2}-E^{2}}}\mathrm{Re}\left[J_{2}\left((k_{F,\lambda}+i\zeta_{\lambda}^{-1})|x|\right)+iH_{-2}\left((k_{F,\lambda}+i\zeta_{\lambda}^{-1})|x|\right)+\frac{2i}{\pi\left(k_{F,\lambda}+i\zeta_{\lambda}^{-1}\right)|x|}\right]\\
K_{2,\lambda}(x;E)&=-\gamma_{\lambda}\mathrm{Im}\left[J_{2}\left((k_{F,\lambda}+i\zeta_{\lambda}^{-1})|x|\right)+iH_{-2}\left((k_{F,\lambda}+i\zeta_{\lambda}^{-1})|x|\right)+\frac{2i}{\pi\left(k_{F,\lambda}+i\zeta_{\lambda}^{-1}\right)|x|}\right]\\
I_{3,\lambda}(x;E)&=\mathrm{sgn}[x]\frac{i\Delta\gamma_{\lambda}}{\sqrt{\Delta^{2}-E^{2}}}\mathrm{Re}\left[J_{3}\left((k_{F,\lambda}+i\zeta_{\lambda}^{-1})|x|\right)-iH_{-3}\left((k_{F,\lambda}+i\zeta_{\lambda}^{-1})|x|\right)+\frac{6i}{\pi\left[\left(k_{F,\lambda}+i\zeta_{\lambda}^{-1}\right)|x|\right]^{2}}\right]\\
K_{3,\lambda}(x;E)&=-\mathrm{sgn}[x]i\gamma_{\lambda}\mathrm{Im}\left[J_{3}\left((k_{F,\lambda}+i\zeta_{\lambda}^{-1})|x|\right)-iH_{-3}\left((k_{F,\lambda}+i\zeta_{\lambda}^{-1})|x|\right)+\frac{6i}{\pi\left[\left(k_{F,\lambda}+i\zeta_{\lambda}^{-1}\right)|x|\right]^{2}}\right]
\end{align}
Here $J_{n}(z)$ and $H_{n}(z)$ are Bessel and Struve functions
of order $n$, respectively; $\zeta_{\lambda}^{-1}\equiv\frac{\sqrt{\Delta^{2}-E^{2}}}{v_{F,\lambda}}$, and
$\gamma_{\lambda}\equiv 1+\lambda\frac{\alpha}{\sqrt{1+\alpha^{2}}}$. Note that the expressions for $K_{l,\lambda}(x;E)$
given above are valid for $x\neq0$, and the integral $K_{l,\lambda}(0;E)=0$ for $x=0$. Assuming $k_{F}|x|\gg1$ and $\zeta_{\lambda}^{-1}\approx\frac{\Delta}{v_{F,\lambda}}\ll k_{F,\lambda}$,
we can use the asymptotic forms of the Bessel and Struve functions~\cite{abramowitz1964}. In the limit $k_{F}x \gg 1$, one can find approximate expressions up to the order $1/(k_{F}x)^{2}$:
\begin{align}
I_{0,\lambda}(x;E)&=\frac{-\Delta\gamma_{\lambda}}{\sqrt{\Delta^{2}-E^{2}}}\sqrt{\frac{2}{\pi k_{F,\lambda}|x|}}e^{-\zeta_{\lambda}^{-1}|x|}\left[\cos(k_{F,\lambda}|x|-\frac{1}{4}\pi)+\frac{1}{8k_{F,\lambda}|x|}\sin(k_{F,\lambda}|x|-\frac{1}{4}\pi)\right]\\
K_{0,\lambda}(x;E)&=\gamma_{\lambda}\sqrt{\frac{2}{\pi k_{F,\lambda}|x|}}e^{-\zeta_{\lambda}^{-1}|x|}\left[\sin(k_{F,\lambda}|x|-\frac{1}{4}\pi)-\frac{1}{8k_{F,\lambda}|x|}\cos(k_{F,\lambda}|x|-\frac{1}{4}\pi)\right]+\frac{2\gamma_{\lambda}}{\pi k_{F,\lambda}|x|}\\
I_{1,\lambda}(x;E)&=-\mathrm{sgn}[x]\frac{i\Delta\gamma_{\lambda}}{\sqrt{\Delta^{2}-E^{2}}}\sqrt{\frac{2}{\pi k_{F,\lambda}|x|}}e^{-\zeta_{\lambda}^{-1}|x|}\left[\cos(k_{F,\lambda}|x|-\frac{3}{4}\pi)-\frac{3}{8k_{F,\lambda}|x|}\sin(k_{F,\lambda}|x|-\frac{3}{4}\pi)\right]\\
K_{1,\lambda}(x;E)&=\mathrm{sgn}[x]i\gamma_{\lambda}\sqrt{\frac{2}{\pi k_{F,\lambda}|x|}}e^{-\zeta_{\lambda}^{-1}|x|}\left[\sin(k_{F,\lambda}|x|-\frac{3}{4}\pi)+\frac{3}{8k_{F,\lambda}|x|}\cos(k_{F,\lambda}|x|-\frac{3}{4}\pi)\right]+\mathrm{sgn}[x]\frac{i2\gamma_{\lambda}}{\pi \left(k_{F,\lambda}|x|\right)^{2}}
\end{align}

\begin{align}
I_{2,\lambda}(x;E)&=\frac{-\Delta\gamma_{\lambda}}{\sqrt{\Delta^{2}-E^{2}}}\sqrt{\frac{2}{\pi k_{F,\lambda}|x|}}e^{-\zeta_{\lambda}^{-1}|x|}\left[\cos(k_{F,\lambda}|x|-\frac{1}{4}\pi)-\frac{15}{8k_{F,\lambda}|x|}\sin(k_{F,\lambda}|x|-\frac{1}{4}\pi)\right]\\
K_{2,\lambda}(x;E)&=\gamma_{\lambda}\sqrt{\frac{2}{\pi k_{F,\lambda}|x|}}e^{-\zeta_{\lambda}^{-1}|x|}\left[\sin(k_{F,\lambda}|x|-\frac{1}{4}\pi)+\frac{15}{8k_{F,\lambda}|x|}\cos(k_{F,\lambda}|x|-\frac{1}{4}\pi)\right]-\frac{2\gamma_{\lambda}}{\pi k_{F,\lambda}|x|}\\
I_{3,\lambda}(x;E)&=-\mathrm{sgn}[x]\frac{i\Delta\gamma_{\lambda}}{\sqrt{\Delta^{2}-E^{2}}}\sqrt{\frac{2}{\pi k_{F,\lambda}|x|}}e^{-\zeta_{\lambda}^{-1}|x|}\left[\cos(k_{F,\lambda}|x|-\frac{3}{4}\pi)-\frac{35}{8k_{F,\lambda}|x|}\sin(k_{F,\lambda}|x|-\frac{3}{4}\pi)\right]\\
K_{3,\lambda}(x;E)&=\mathrm{sgn}[x]i\gamma_{\lambda}\sqrt{\frac{2}{\pi k_{F,\lambda}|x|}}e^{-\zeta_{\lambda}^{-1}|x|}\left[\sin(k_{F,\lambda}|x|-\frac{3}{4}\pi)+\frac{35}{8k_{F,\lambda}|x|}\cos(k_{F,\lambda}|x|-\frac{3}{4}\pi)\right]-\mathrm{sgn}[x]\frac{i6\gamma_{\lambda}}{\pi \left(k_{F,\lambda}|x|\right)^{2}}
\end{align}

The corresponding Fourier transforms of the above asymptotic forms to the leading order of $\frac{1}{\sqrt{k_{F}a}}$ are given by
\begin{align}\label{eq:I0}
I_{0,\lambda}(k;E)=I_{2,\lambda}(k;E) & =\frac{-\Delta\gamma_{\lambda}}{\sqrt{\Delta^{2}-E^{2}}}\sqrt{\frac{1}{2\pi k_{F,\lambda}a}}\Biggl[e^{-i\frac{1}{4}\pi}\mathrm{Li}_{\frac{1}{2}}\left(e^{ik_{F,\lambda}a-\zeta_{\lambda}^{-1}a+ika}\right)+e^{i\frac{1}{4}\pi}\mathrm{Li}_{\frac{1}{2}}\left(e^{-ik_{F,\lambda}a-\zeta_{\lambda}^{-1}a+ika}\right)\nonumber \\
 & \ \ \ \ \ \ \ \ \ \ \ \ \ \ \ \ \ \ \ \ \ \ \ \ \ +e^{-i\frac{1}{4}\pi}\mathrm{Li}_{\frac{1}{2}}\left(e^{ik_{F,\lambda}a-\zeta_{\lambda}^{-1}a-ika}\right)+e^{i\frac{1}{4}\pi}\mathrm{Li}_{\frac{1}{2}}\left(e^{-ik_{F,\lambda}a-\zeta_{\lambda}^{-1}a-ika}\right)\Biggr]
\end{align}
\begin{align}\label{eq:K0}
K_{0,\lambda}(k;E)=K_{2,\lambda}(k;E) & =-i\gamma_{\lambda}\sqrt{\frac{1}{2\pi k_{F,\lambda}a}}\Biggl[e^{-i\frac{1}{4}\pi}\mathrm{Li}_{\frac{1}{2}}\left(e^{ik_{F,\lambda}a-\zeta_{\lambda}^{-1}a+ika}\right)-e^{i\frac{1}{4}\pi}\mathrm{Li}_{\frac{1}{2}}\left(e^{-ik_{F,\lambda}a-\zeta_{\lambda}^{-1}a+ika}\right)\nonumber \\
 & \ \ \ \ \ \ \ \ \ \ \ \ \ \ \ \ \ \ \ \ \ \ \ +e^{-i\frac{1}{4}\pi}\mathrm{Li}_{\frac{1}{2}}\left(e^{ik_{F,\lambda}a-\zeta_{\lambda}^{-1}a-ika}\right)-e^{i\frac{1}{4}\pi}\mathrm{Li}_{\frac{1}{2}}\left(e^{-ik_{F,\lambda}a-\zeta_{\lambda}^{-1}a-ika}\right)\Biggr]
\end{align}
\begin{align}\label{eq:I1}
I_{1,\lambda}(k;E)=I_{3,\lambda}(k;E) & =-\frac{i\Delta\gamma_{\lambda}}{\sqrt{\Delta^{2}-E^{2}}}\sqrt{\frac{1}{2\pi k_{F,\lambda}a}}\Biggl[e^{-i\frac{3}{4}\pi}\mathrm{Li}_{\frac{1}{2}}\left(e^{ik_{F,\lambda}a-\zeta_{\lambda}^{-1}a+ika}\right)+e^{i\frac{3}{4}\pi}\mathrm{Li}_{\frac{1}{2}}\left(e^{-ik_{F,\lambda}a-\zeta_{\lambda}^{-1}a+ika}\right)\nonumber \\
 & \ \ \ \ \ \ \ \ \ \ \ \ \ \ \ \ \ \ \ \ \ \ \ \ \ -e^{-i\frac{3}{4}\pi}\mathrm{Li}_{\frac{1}{2}}\left(e^{ik_{F,\lambda}a-\zeta_{\lambda}^{-1}a-ika}\right)-e^{i\frac{3}{4}\pi}\mathrm{Li}_{\frac{1}{2}}\left(e^{-ik_{F,\lambda}a-\zeta_{\lambda}^{-1}a-ika}\right)\Biggr]
\end{align}
\begin{align}\label{eq:K1}
K_{1,\lambda}(k;E)=K_{3,\lambda}(k;E) & =\gamma_{\lambda}\sqrt{\frac{1}{2\pi k_{F,\lambda}a}}\Biggl[e^{-i\frac{3}{4}\pi}\mathrm{Li}_{\frac{1}{2}}\left(e^{ik_{F,\lambda}a-\zeta_{\lambda}^{-1}a+ika}\right)-e^{i\frac{3}{4}\pi}\mathrm{Li}_{\frac{1}{2}}\left(e^{-ik_{F,\lambda}a-\zeta_{\lambda}^{-1}a+ika}\right)\nonumber \\
 & \ \ \ \ \ \ \ \ \ \ \ \ \ \ \ \ \ \ \ \ \ \ \ -e^{-i\frac{3}{4}\pi}\mathrm{Li}_{\frac{1}{2}}\left(e^{ik_{F,\lambda}a-\zeta_{\lambda}^{-1}a-ika}\right)+e^{i\frac{3}{4}\pi}\mathrm{Li}_{\frac{1}{2}}\left(e^{-ik_{F,\lambda}a-\zeta_{\lambda}^{-1}a-ika}\right)\Biggr]
\end{align}
where $\mathrm{Li}_{s}(z)$ is the polylogarithm function
\[
\mathrm{Li}_{s}(z)=\sum_{n=1}^{\infty}\frac{z^{n}}{n^{s}}.
\]

\section{Calculation of the Greens functions $\mathcal{G}_{l-l'}^{ij}(E)$}\label{app:B}

The Green's function of a superconductor with Rashba spin-orbit coupling can be expanded in the angular momentum channels $G(\mathbf{k};E)=\sum_{l}G_{l}(k;E)e^{il\theta_{\mathbf{k}}}$. The local Green's function (i.e. $i=j$) reads
\begin{equation}
\mathcal{G}_{l-l'}^{ii}(E)=\int\frac{kdk}{2\pi}G_{l-l'}(k;E)\equiv\overline{G}_{l-l'}(E).
\end{equation}
Using Eqs.~(\ref{eq:gf1})-(\ref{eq:gf3}), one finds that there are three non-zero local Green's function corresponding to angular momenta $l=-1,\,0,\,1$:
\begin{align}
\overline{G}_{-1}(E) & =\frac{i\pi\left(N_{+}-N_{-}\right)}{2\sqrt{\Delta^{2}-E^{2}}}\left(E\sigma_{+}\tau_{0}+\Delta\sigma_{+}\tau_{x}\right),\\
\overline{G}_{0}(E) & =-\frac{\pi\left(N_{+}+N_{-}\right)}{2\sqrt{\Delta^{2}-E^{2}}}\left(E\sigma_{0}\tau_{0}+\Delta\sigma_{0}\tau_{x}\right),\\
\overline{G}_{1}(E) & =-\frac{i\pi\left(N_{+}-N_{-}\right)}{2\sqrt{\Delta^{2}-E^{2}}}\left(E\sigma_{-}\tau_{0}+\Delta\sigma_{-}\tau_{x}\right).
\end{align}
The non-local Green's function ($i\neq j$) is defined as
\begin{align}
\mathcal{G}_{m}^{ij}(E)=\frac{\pi N_{F}}{2}\sum_{\lambda=\pm} & \Biggl\{\left(-i\lambda\right)\left[\left(\frac{E}{\Delta}\sigma_{+}\tau_{0}+\sigma_{+}\tau_{x}\right)I_{\left|-1+m\right|,\lambda}+\left(\sigma_{+}\tau_{z}\right)K_{\left|-1+m\right|,\lambda}\right]\nonumber \\
 & +\left[\left(\frac{E}{\Delta}\sigma_{0}\tau_{0}+\sigma_{0}\tau_{x}\right)I_{\left|m\right|,\lambda}+\left(\sigma_{0}\tau_{z}\right)K_{\left|m\right|,\lambda}\right]\nonumber \\
 & +\left(i\lambda\right)\left[\left(\frac{E}{\Delta}\sigma_{-}\tau_{0}+\sigma_{-}\tau_{x}\right)I_{\left|1+m\right|,\lambda}+\left(\sigma_{-}\tau_{z}\right)K_{\left|1+m\right|,\lambda}\right]\Biggr\}
\end{align}
where $m=l-l'=0,\,\pm1,\,\pm2$, and the functions $I_{n,\lambda}(x_{ij};E)$ and $K_{n,\lambda}(x_{ij};E)$ are given in Appendix~\ref{app:A}.

\section{Derivation of effective Hamiltonian in deep $s$-band limit}\label{app:C}

In this Appendix we provide the details of the derivation of effective Hamiltonian in deep $s$-band limit. Using the method outlined in Sec.~\ref{sec:2}, we find that the effective eigenvalue equation for the deep $s$-band limit reads
\begin{equation}
\sum_{j,l}\mathbf{M}_{0,l}^{ij}(E)\overline{\psi}_{j,l}=0.
\end{equation}
After substituting Eq.~(\ref{eq:s2}) back into the above equation, one finds
\begin{align}
0= & (\mathbf{M}_{0,0}^{ii}-\mathbf{M}_{0,-1}^{ii}(\mathbf{M}_{-1,-1}^{ii})^{-1}\mathbf{M}_{-1,0}^{ii}-\mathbf{M}_{0,1}^{ii}(\mathbf{M}_{1,1}^{ii})^{-1}\mathbf{M}_{1,0}^{ii})\overline{\psi}_{i,0}+\sum_{j\neq i}(\mathbf{M}_{0,-1}^{ij}\overline{\psi}_{j,-1}+\mathbf{M}_{0,0}^{ij}\overline{\psi}_{j,0}+\mathbf{M}_{0,1}^{ij}\overline{\psi}_{j,1})\nonumber \\
 & +\mathbf{M}_{0,-1}^{ii}(\mathbf{M}_{-1,-1}^{ii})^{-1}\sum_{j\neq i}(\mathbf{M}_{-1,-1}^{ij}\overline{\psi}_{j,-1}+\mathbf{M}_{-1,0}^{ij}\overline{\psi}_{j,0}+\mathbf{M}_{-1,1}^{ij}\overline{\psi}_{j,1})\\
 & +\mathbf{M}_{0,1}^{ii}(\mathbf{M}_{1,1}^{ii})^{-1}\sum_{j\neq i}(\mathbf{M}_{1,-1}^{ij}\overline{\psi}_{j,-1}+\mathbf{M}_{1,0}^{ij}\overline{\psi}_{j,0}+\mathbf{M}_{1,1}^{ij}\overline{\psi}_{j,1}).\nonumber
\end{align}
The assumption $k_F a \gg 1$ allows one to neglect the terms $\mathcal{O}((\mathbf{M}^{i\neq j})^{2})$. Using the tight-binding approximation, one finally arrives at
\begin{align}
0= & \bigg[\mathbf{M}_{0,0}^{ii}-\mathbf{M}_{0,-1}^{ii}(\mathbf{M}_{-1,-1}^{ii})^{-1}\mathbf{M}_{-1,0}^{ii}-\mathbf{M}_{0,1}^{ii}(\mathbf{M}_{1,1}^{ii})^{-1}\mathbf{M}_{1,0}^{ii}\bigg]\overline{\psi}_{i,0}\nonumber \\
 & -\sum_{j\neq i}\bigg[\mathbf{M}_{0,0}^{ij}-\mathbf{M}_{0,-1}^{ij}(\mathbf{M}_{-1,-1}^{ii})^{-1}\mathbf{M}_{-1,0}^{ii}-\mathbf{M}_{0,1}^{ij}(\mathbf{M}_{1,1}^{ii})^{-1}\mathbf{M}_{1,0}^{ii}-\mathbf{M}_{0,-1}^{ii}(\mathbf{M}_{-1,-1}^{ii})^{-1}\mathbf{M}_{-1,0}^{ij}-\mathbf{M}_{0,1}^{ii}(\mathbf{M}_{1,1}^{ii})^{-1}\mathbf{M}_{1,0}^{ij}\bigg]\overline{\psi}_{j,0}\nonumber \\
 & -\sum_{j\neq i}\bigg[\mathbf{M}_{0,-1}^{ii}(\mathbf{M}_{-1,-1}^{ii})^{-1}\mathbf{M}_{-1,-1}^{ij}(\mathbf{M}_{-1,-1}^{ii})^{-1}\mathbf{M}_{-1,0}^{ii}+\mathbf{M}_{0,-1}^{ii}(\mathbf{M}_{-1,-1}^{ii})^{-1}\mathbf{M}_{-1,1}^{ij}(\mathbf{M}_{1,1}^{ii})^{-1}\mathbf{M}_{1,0}^{ii}\nonumber \\
 & +\mathbf{M}_{0,1}^{ii}(\mathbf{M}_{1,1}^{ii})^{-1}\mathbf{M}_{1,-1}^{ij}(\mathbf{M}_{-1,-1}^{ii})^{-1}\mathbf{M}_{-1,0}^{ii}+\mathbf{M}_{0,1}^{ii}(\mathbf{M}_{1,1}^{ii})^{-1}\mathbf{M}_{1,1}^{ij}(\mathbf{M}_{1,1}^{ii})^{-1}\mathbf{M}_{1,0}^{ii}\bigg]\overline{\psi}_{j,0}+\mathcal{O}((\mathbf{M}^{ij})^{2})\nonumber\\
\equiv & \sum_{j}\mathbf{M}_{s}^{ij}(E)\overline{\psi}_{j,0}\label{eq:l0band}
\end{align}

\section{Local basis in the deep $s$-band limit}\label{app:D}

In this Appendix, we provide details for the projection procedure used to the derive an effective Hamiltonian for the deep $s$-band limit.
Assuming that the magnetic-atom polarization is along $\hat{\mathbf{z}}$ axis, the unnormalized local
basis for magnetic impurity reads:
$$\varphi_{+}\sim\left(\begin{array}{cccc}
1, & 0, & 1, & 0\end{array}\right)^{T} \mbox{ and }\varphi_{-}\sim\left(\begin{array}{cccc}
0, & 1, & 0, & -1\end{array}\right)^{T}.$$
When the polarization is along $\hat{\mathbf{x}}$-axis, the unnormalized local
basis for magnetic impurity is given by
$$\varphi_{+}\sim\left(\begin{array}{cccc}
1, & 1, & 1, & 1\end{array}\right)^{T} \mbox{ and } \varphi_{-}\sim\left(\begin{array}{cccc}
-1, & 1, & 1, & -1\end{array}\right)^{T}.$$

For the sake of completeness, we also derive effective Hamiltonian when the magnetization is along $y$-axis. In this case, the unnormalized local
basis for magnetic impurity becomes
$$\varphi_{+}\sim\left(\begin{array}{cccc}
1, & i, & 1, & i\end{array}\right)^{T} \mbox{ and } \varphi_{-}\sim\left(\begin{array}{cccc}
i, & 1, & -i, & -1\end{array}\right)^{T}.$$
After the projection onto the local basis, the effective Hamiltonian is given by
\begin{equation}
\frac{\mathcal{H}_{s}^{\hat{y}}(k)}{\Delta}=\left(\begin{array}{cc}
h_{y}(k)+d_{y}(k) & 0\\
0 & -h_{y}(k)+d_{y}(k)
\end{array}\right),
\end{equation}
with the functions $h_{y}(k)$ and $d_{y}(k)$ being
\begin{align}
h_{y}(k) & =\epsilon_{y}+\frac{1}{2}\left[I_{0,+}(k)+I_{0,-}(k)\right],\\
d_{y}(k) & =\frac{1}{2}\left[I_{1,+}(k)-I_{1,-}(k)\right]+\frac{J_{1}\alpha}{\left(1-J_{1}\right)}\left[I_{1,+}(k)+I_{1,-}(k)\right].
\end{align}
Here the on-site energy $\epsilon_{y}=\epsilon_{x}$. One can see that quasiparticle spectrum in this case is indeed gapless.

\section{Effective Hamiltonian in the long wavelength limit $k\rightarrow 0$}\label{app:G}
It is instructive to expand the functions $I(n,k)$ and $K(n,k)$ appearing in our effective Hamiltonian close to $k=0$ in order to understand the spectrum qualitatively. After some algebra, one finds
\ba
I_{0,\lambda}(k;E=0)&=&I_{2,\lambda}(k;E=0)=-2\gamma_{\lambda}\sqrt{\frac{1}{\pi k_{F,\lambda}a}}A_{0}(k_{F,\lambda}a+i\zeta_\lambda^{-1}a)-2\gamma_{\lambda}\sqrt{\frac{1}{\pi k_{F,\lambda}a}}A_{2}(k_{F,\lambda}a+i\zeta_\lambda^{-1}a)k^2+\mathcal{O}(k^4),\\
I_{1,\lambda}(k;E=0)&=&I_{3,\lambda}(k;E=0)=-2\gamma_{\lambda}\sqrt{\frac{1}{\pi k_{F,\lambda}a}}B_{1}(k_{F,\lambda}a+i\zeta_\lambda^{-1}a)k+\mathcal{O}(k^3),\\
K_{1,\lambda}(k;E=0)&=&K_{3,\lambda}(k;E=0)=2\gamma_{\lambda}\sqrt{\frac{1}{\pi k_{F,\lambda}a}}C_{1}(k_{F,\lambda}a+i\zeta_\lambda^{-1}a)k+\mathcal{O}(k^3),
\ea
where
\ba
A_{0}(z)&=&\mathrm{Re}\left[\mathrm{Li}_{\frac{1}{2}}\left(e^{iz}\right)\right] +\mathrm{Im}\left[\mathrm{Li}_{\frac{1}{2}}\left(e^{iz}\right)\right] ,\\
A_{2}(z)&=&\mathrm{Im}\left[\mathrm{Li}_{-\frac{3}{2}}\left(e^{iz}\right)\right],\\
B_{1}(z)&=&\mathrm{Re}\left[\mathrm{Li}_{-\frac{1}{2}}\left(e^{iz}\right)\right]-\mathrm{Im}\left[\mathrm{Li}_{-\frac{1}{2}}\left(e^{iz}\right)\right] ,\\
C_{1}(z)&=&\mathrm{Re}\left[\mathrm{Li}_{-\frac{1}{2}}\left(e^{iz}\right)\right]+\mathrm{Im}\left[\mathrm{Li}_{-\frac{1}{2}}\left(e^{iz}\right)\right] .
\ea

The dependence of the functions $A_{0}(z)$, $A_{2}(z)$, $B_{1}(z)$ and $C_{1}(z)$ on the external parameters is shown in Fig.~\ref{fig:g1}. One can notice that when $k_{F,\lambda}a=2\pi n$ with $n$ being an integer, these functions have singularities which follows from the definition of polylogarithm function. These singularities are cutoff by the finite coherence length. In realistic systems, however, the superconducting coherence length is much larger than the interatomic spacing, and, thus, the parameters such as effective mass and Fermi velocity are strongly dependent on $k_F a$, see Fig.~\ref{fig:g1}.
\begin{figure}[H]
\includegraphics[height=1.1in]{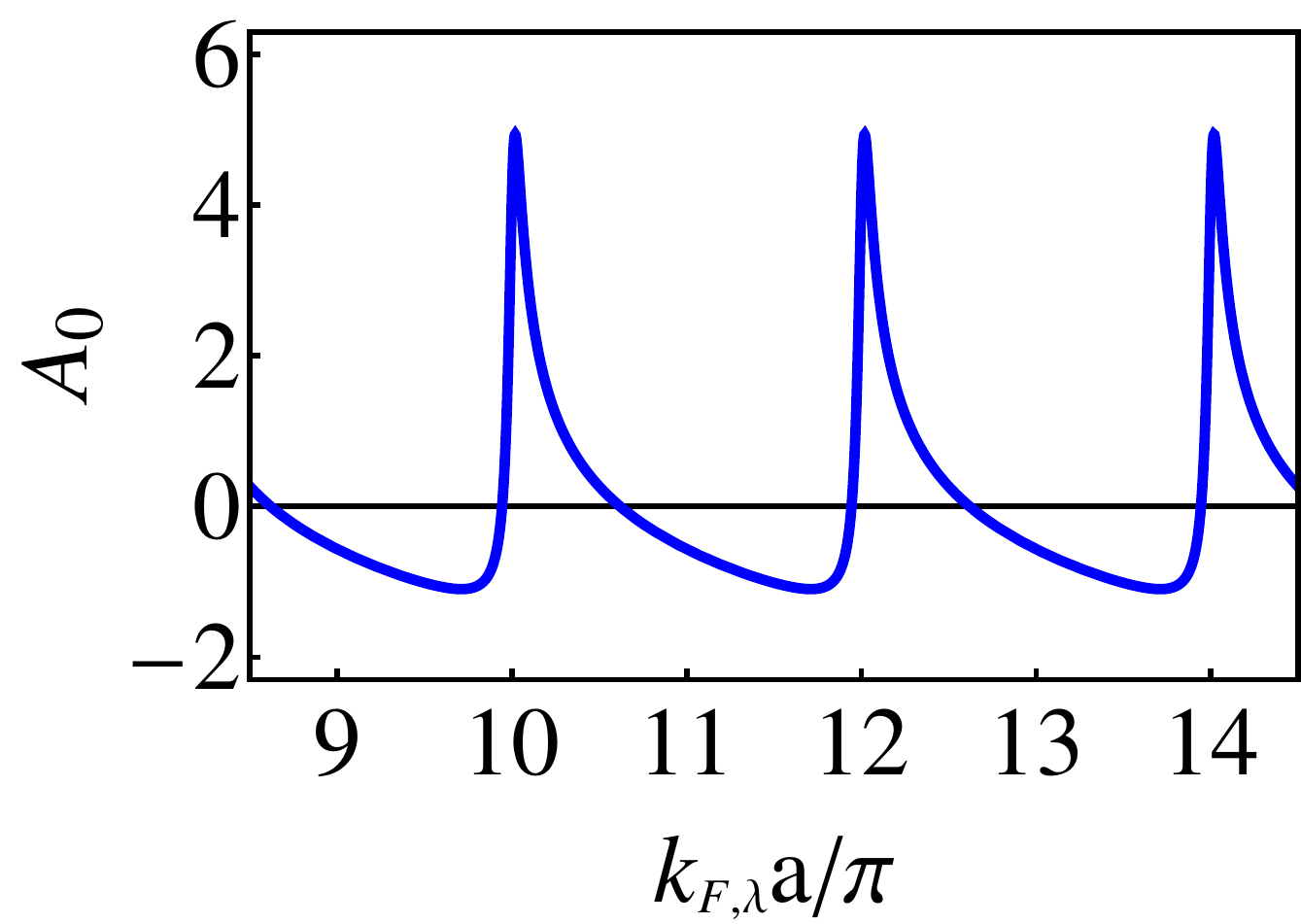}\,\,\,\,\,\includegraphics[height=1.1in]{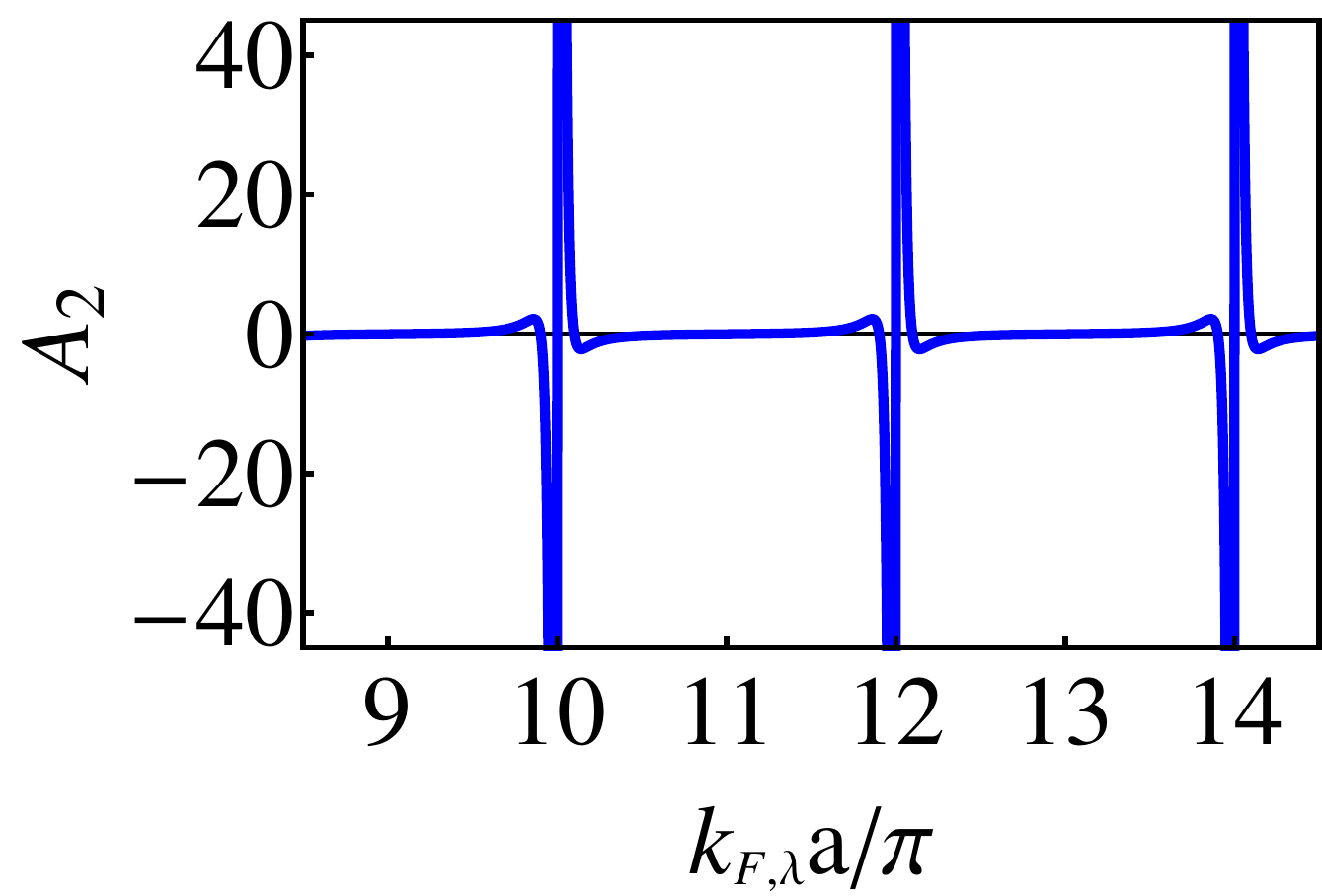}\,\,\,\,\,\includegraphics[height=1.1in]{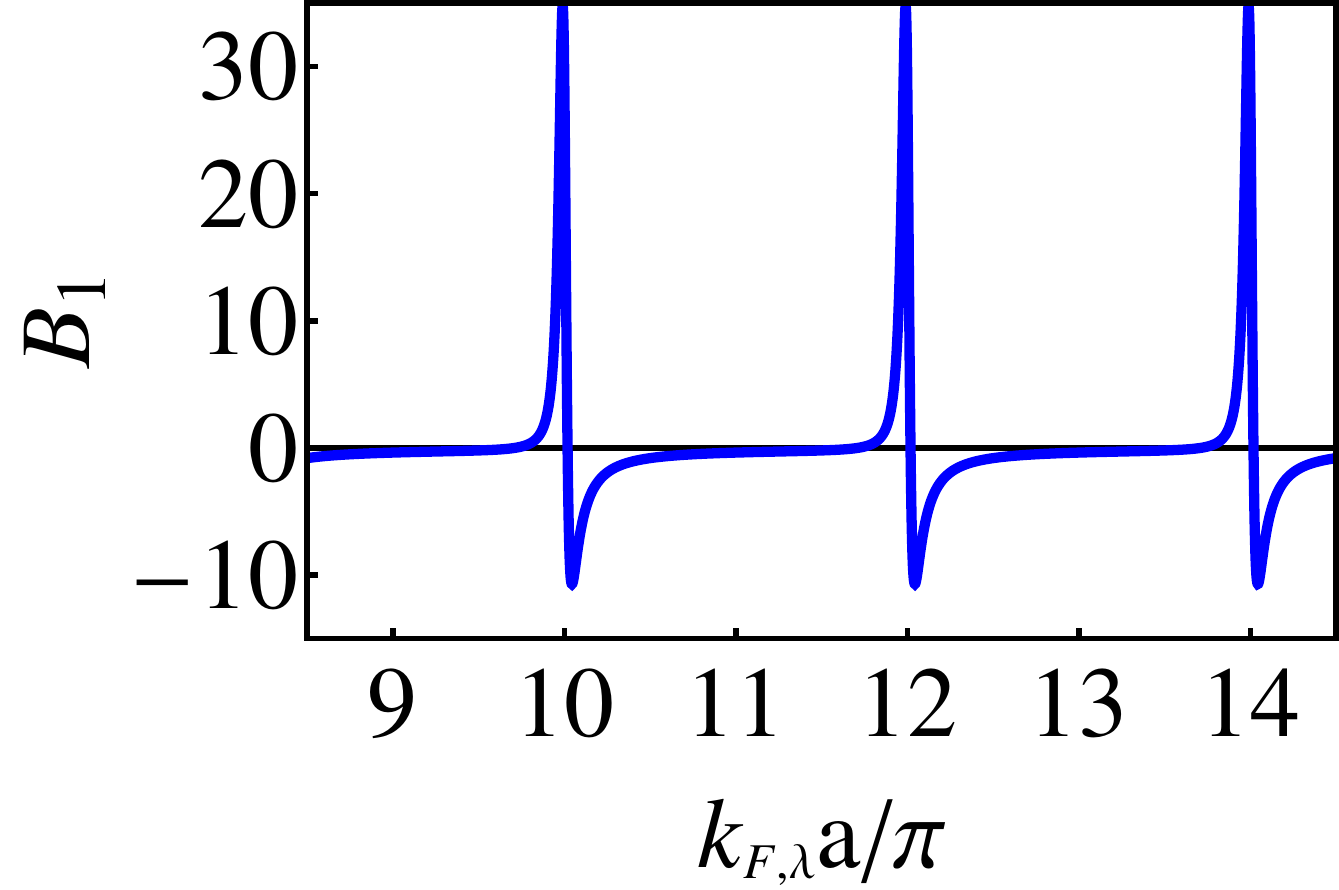}\,\,\,\,\,\includegraphics[height=1.1in]{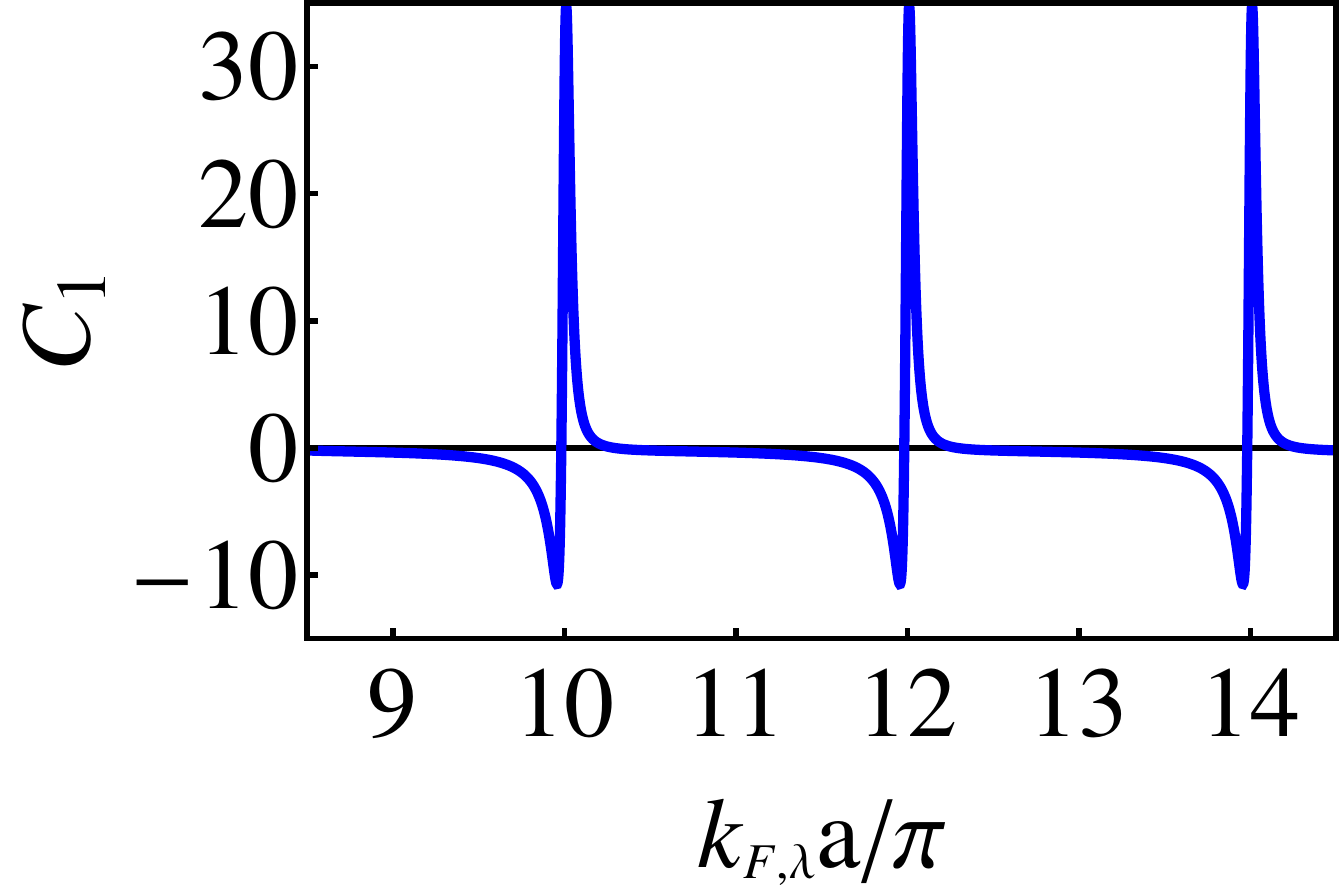}
\caption{(Color online) The dependence of the functions $A_{0}(k_{F,\lambda}a+i\zeta_\lambda^{-1}a)$, $A_{2}(k_{F,\lambda}a+i\zeta_\lambda^{-1}a)$, $B_{1}(k_{F,\lambda}a+i\zeta_\lambda^{-1}a)$ and $C_{1}(k_{F,\lambda}a+i\zeta_\lambda^{-1}a)$ on $k_{F,\lambda}a$. Here we used  $\zeta_\lambda=10a$. \label{fig:g1}}
\end{figure}

Finally, the expansion of the coefficients in the deep $s$-band Hamiltonian at $k\rightarrow 0$ becomes
\ba
h_z^{(0)}&\!=\!&\epsilon_z-\sum_\lambda \gamma_{\lambda} \sqrt{\frac{1}{\pi k_{F,\lambda}a}} \left(1+\frac{2\alpha J_1(\alpha-\lambda)}{1+J_1}\right)A_{0}(k_{F,\lambda}a+i\zeta_\lambda^{-1}a),\\
h_z^{(2)}&\!=\!&-\sum_\lambda \gamma_{\lambda} \sqrt{\frac{1}{\pi k_{F,\lambda}a}}\left(1+\frac{2\alpha J_1(\alpha-\lambda)}{1+J_1}\right)A_{2}(k_{F,\lambda}a+i\zeta_\lambda^{-1}a),\\
h_x^{(0)}&\!=\!&\epsilon_x\!-\sum_\lambda \gamma_{\lambda} \sqrt{\frac{1}{\pi k_{F,\lambda}a}} \left(1\!-\!\frac{2\lambda\alpha J_1}{1-J_1}\right)A_{0}(k_{F,\lambda}a+i\zeta_\lambda^{-1}a),\\
h_x^{(2)}&\!=\!&-\sum_\lambda \gamma_{\lambda} \sqrt{\frac{1}{\pi k_{F,\lambda}a}}\left(1\!-\!\frac{2\lambda\alpha J_1}{1-J_1}\!\right)A_{2}(k_{F,\lambda}a+i\zeta_\lambda^{-1}a),\\
h_y^{(0)}&\!=\!&\epsilon_x\!-\sum_\lambda \gamma_{\lambda} \sqrt{\frac{1}{\pi k_{F,\lambda}a}}\left(1\!+\!\!\frac{2\lambda\alpha J_1}{1-J_1}\!\right)A_{0}(k_{F,\lambda}a+i\zeta_\lambda^{-1}a),\\
h_y^{(2)}&\!=\!&-\sum_\lambda \gamma_{\lambda} \sqrt{\frac{1}{\pi k_{F,\lambda}a}}\left(1\!+\!\frac{2\lambda\alpha J_1}{1-J_1}\!\right)A_{2}(k_{F,\lambda}a+i\zeta_\lambda^{-1}a),\\
\Delta^{(1)}&\!=\!&\sum_\lambda i \gamma_{\lambda} \sqrt{\frac{1}{\pi k_{F,\lambda}a}}\left(\lambda -\frac{2\alpha J_1}{1+J_1} \right)C_{1}(k_{F,\lambda}a+i\zeta_\lambda^{-1}a),\\
d_y^{(1)}&\!=\!&-\sum_\lambda \gamma_{\lambda} \sqrt{\frac{1}{\pi k_{F,\lambda}a}}\left(\lambda +\frac{2\alpha J_1}{1-J_1} \right)B_{1}(k_{F,\lambda}a+i\zeta_\lambda^{-1}a).
\ea
The expansion coefficients in the deep $p$-band Hamiltonian are
\ba
h_{11}^{(0)}&=&\epsilon_1-\sum_\lambda \gamma_{\lambda}\sqrt{\frac{1}{\pi k_{F,\lambda}a}}A_{0}(k_{F,\lambda}a+i\zeta_\lambda^{-1}a),\\
h_{22}^{(0)}&=&\epsilon_2-\sum_\lambda \gamma_{\lambda}\sqrt{\frac{1}{\pi k_{F,\lambda}a}}A_{0}(k_{F,\lambda}a+i\zeta_\lambda^{-1}a),\\
h_{12}^{(0)}&=&=-\sum_\lambda \gamma_{\lambda}\sqrt{\frac{1}{\pi k_{F,\lambda}a}}A_{0}(k_{F,\lambda}a+i\zeta_\lambda^{-1}a),\\
h_{11}^{(2)}&=&h_{22}^{(2)}=h_{12}^{(2)}=-\sum_\lambda \gamma_{\lambda}\sqrt{\frac{1}{\pi k_{F,\lambda}a}}A_{2}(k_{F,\lambda}a+i\zeta_\lambda^{-1}a),
\ea
\ba
\Delta_{11}^{(1)}&=&\sum_\lambda i \gamma_{\lambda} \sqrt{\frac{1}{\pi k_{F,\lambda}a}}\left(\lambda-\frac{2\alpha J_0}{1+J_0}\right)C_{1}(k_{F,\lambda}a+i\zeta_\lambda^{-1}a),\\
\Delta_{22}^{(1)}&=&\sum_\lambda i \lambda \gamma_{\lambda}\sqrt{\frac{1}{\pi k_{F,\lambda}a}}C_{1}(k_{F,\lambda}a+i\zeta_\lambda^{-1}a),\\
\Delta_{12}^{(1)}&=&\sum_\lambda i \gamma_{\lambda} \sqrt{\frac{1}{\pi k_{F,\lambda}a}}\left(\lambda-\frac{\alpha J_0}{1+J_0}\right)C_{1,}(k_{F,\lambda}a+i\zeta_\lambda^{-1}a).
\ea

\section{Derivation of effective Hamiltonian in deep p-band limit}\label{app:E}

In this Appendix, we provide the details of the derivation of effective Hamiltonian in deep p-band limit. Using the method outlined in Sec.~\ref{sec:2}, we find the effective eigenvalue equation for the deep $p$-band limit. The corresponding equations for p-wave bands are given by
\begin{eqnarray}
\sum_{j,l}\mathbf{M}_{-1,l}^{ij}(E)\overline{\psi}_{j,l} & = & 0\label{eq:p1}\\
\sum_{j,l}\mathbf{M}_{1,l}^{ij}(E)\overline{\psi}_{j,l} & = & 0\label{eq:p2}
\end{eqnarray}
In order to integrate out s-channel, we have to solve for $\overline{\psi}_{i,0}$ finding that
\begin{equation}
\overline{\psi}_{i,0}=-(\mathbf{M}_{1,0}^{ii})^{-1}(\mathbf{M}_{0,-1}^{ii}\overline{\psi}_{i,-1}+\mathbf{M}_{0,1}^{ii}\overline{\psi}_{i,1}+\sum_{j\neq i,l}\mathbf{M}_{0,l}^{ij}\overline{\psi}_{j,l}).\label{eq:p3}
\end{equation}
Substituting Eq.~(\ref{eq:p3}) into Eq.~(\ref{eq:p1}, \ref{eq:p2})
and following the same procedure as in Appendix \ref{app:C}, we eventually obtain two coupled equations for the p-wave bands
\begin{align}
0= & (\mathbf{M}_{-1,-1}^{ii}-\mathbf{M}_{-1,0}^{ii}(\mathbf{M}_{0,0}^{ii})^{-1}\mathbf{M}_{0,-1}^{ii})\overline{\psi}_{i,-1}-\mathbf{M}_{-1,0}^{ii}(\mathbf{M}_{0,0}^{ii})^{-1}\mathbf{M}_{0,1}^{ii}\overline{\psi}_{i,1}\nonumber \\
 & -\sum_{j\neq i}\bigg[\mathbf{M}_{-1,-1}^{ij}-\mathbf{M}_{-1,0}^{ij}(\mathbf{M}_{0,0}^{ii})^{-1}\mathbf{M}_{0,-1}^{ii}-\mathbf{M}_{-1,0}^{ii}(\mathbf{M}_{0,0}^{ii})^{-1}\mathbf{M}_{0,-1}^{ij}+\mathbf{M}_{-1,0}^{ii}(\mathbf{M}_{0,0}^{ii})^{-1}\mathbf{M}_{0,0}^{ij}(\mathbf{M}_{0,0}^{ii})^{-1}\mathbf{M}_{0,-1}^{ii}\bigg]\overline{\psi}_{j,-1}\nonumber \\
 & -\sum_{j\neq i}\bigg[\mathbf{M}_{-1,1}^{ij}-\mathbf{M}_{-1,0}^{ij}(\mathbf{M}_{0,0}^{ii})^{-1}\mathbf{M}_{0,1}^{ii}-\mathbf{M}_{-1,0}^{ii}(\mathbf{M}_{0,0}^{ii})^{-1}\mathbf{M}_{0,1}^{ij}+\mathbf{M}_{-1,0}^{ii}(\mathbf{M}_{0,0}^{ii})^{-1}\mathbf{M}_{0,0}^{ij}(\mathbf{M}_{0,0}^{ii})^{-1}\mathbf{M}_{0,1}^{ii})\bigg]\overline{\psi}_{j,1}\\
0= & -\mathbf{M}_{1,0}^{ii}(\mathbf{M}_{0,0}^{ii})^{-1}\mathbf{M}_{0,-1}^{ii}\overline{\psi}_{i,-1}+(\mathbf{M}_{1,1}^{ii}-\mathbf{M}_{1,0}^{ii}(\mathbf{M}_{0,0}^{ii})^{-1}\mathbf{M}_{0,1}^{ii})\overline{\psi}_{i,1}\nonumber \\
 & -\sum_{j\neq i}\bigg[\mathbf{M}_{1,-1}^{ij}-\mathbf{M}_{1,0}^{ij}(\mathbf{M}_{0,0}^{ii})^{-1}\mathbf{M}_{0,-1}^{ii}-\mathbf{M}_{1,0}^{ii}(\mathbf{M}_{0,0}^{ii})^{-1}\mathbf{M}_{0,-1}^{ij}+\mathbf{M}_{1,0}^{ii}(\mathbf{M}_{0,0}^{ii})^{-1}\mathbf{M}_{0,0}^{ij}(\mathbf{M}_{0,0}^{ii})^{-1}\mathbf{M}_{0,-1}^{ii})\bigg]\overline{\psi}_{j,-1}\nonumber \\
 & -\sum_{j\neq i}\bigg[\mathbf{M}_{1,1}^{ij}-\mathbf{M}_{1,0}^{ij}(\mathbf{M}_{0,0}^{ii})^{-1}\mathbf{M}_{0,1}^{ii}-\mathbf{M}_{1,0}^{ii}(\mathbf{M}_{0,0}^{ii})^{-1}\mathbf{M}_{0,1}^{ij}+\mathbf{M}_{1,0}^{ii}(\mathbf{M}_{0,0}^{ii})^{-1}\mathbf{M}_{0,0}^{ij}(\mathbf{M}_{0,0}^{ii})^{-1}\mathbf{M}_{0,1}^{ii}\bigg]\overline{\psi}_{j,1}
\end{align}
After some manipulations, one can write eigenvalue equations in the compact form, see Eq.~(\ref{eq: p_channel_chain_Eq}).

\section{Local basis in the p-band system}\label{app:F}

In this Appendix, we discuss the projection procedure in the $p$-band limit in the case of magnetic impurity spins being aligned along $\hat{\mathbf{z}}$ axis. In this case, the unnormalized local spinors are given by
\[
\phi_{1,+}\sim\left(\begin{array}{cccccccc}
1, & 0, & 1, & 0, & 0, & 0, & 0, & 0\end{array}\right)^{T},\ \ \ \phi_{2,+}\sim\left(\begin{array}{cccccccc}
0, & 0, & 0, & 0, & 1, & 0, & 1, & 0\end{array}\right)^{T};
\]

\[
\phi_{1,-}\sim\left(\begin{array}{cccccccc}
0, & 0, & 0, & 0, & 0, & 1, & 0, & -1\end{array}\right)^{T},\ \ \ \phi_{2,-}\sim\left(\begin{array}{cccccccc}
0, & 1, & 0, & -1, & 0, & 0, & 0, & 0\end{array}\right)^{T}.
\]
Projecting onto the local basis $\left(\begin{array}{cccc}
\phi_{1,+}, & \phi_{2,+}, & \phi_{1,-}, & \phi_{2,-}\end{array}\right)^{T}$, one arrives at the two-band Hamiltonian defined in Eq.\,(\ref{eq: p_band_H})-(\ref{eq: inter_pairing_p_band}).

\end{widetext}


%

\end{document}